\documentclass[review]{elsarticle}

\usepackage[hyphens]{url}
\usepackage{lineno,hyperref}
\usepackage{graphicx}
\usepackage{textcomp}
\usepackage{booktabs}
\usepackage{amsmath,amsfonts,amssymb,commath}
\usepackage{subfigure}
\usepackage{multirow}
\usepackage{multicol}
\usepackage{csvsimple}
\usepackage[ruled,vlined,linesnumbered,titlenumbered]{algorithm2e}

\newcommand{\R}{\mathbb{R}}

\journal{}

\begin{document}

\begin{frontmatter}

\title{Real-time Electrical Power Prediction in a Combined Cycle Power Plant}

\author[label1]{Jesus L. Lobo\corref{cor1}}
\author[label2]{Igor Ballesteros} 
\author[label1]{Izaskun Oregi} 
\author[label1,label2,label3]{Javier Del Ser} 

\address[label1]{TECNALIA, Parque Cient\'{\i}fico y Tecnol\'{o}gico de Bizkaia, Astondo Bidea, Edificio 700. E-48160 Derio (Bizkaia), Spain.}
\address[label2]{University of the Basque Country UPV/EHU, 48013 Bilbao, Spain}
\address[label3]{Basque Center for Applied Mathematics (BCAM), 48009 Bilbao, Spain}
\cortext[cor1]{Corresponding author: jesus.lopez@tecnalia.com (Jesus L. Lobo). TECNALIA, Parque Tecnol\'{o}gico de Bizkaia, Edificio 700, 48160 Derio, Spain}

\begin{abstract}
The prediction of electrical power in combined cycle power plants is a key challenge in the electrical power and energy systems field. This power output can vary depending on environmental variables, such as temperature, pressure, and humidity. Thus, the business problem is how to predict the power output as a function of these environmental conditions in order to maximize the profit. The research community has solved this problem by applying machine learning techniques and has managed to reduce the computational and time costs in comparison with the traditional thermodynamical analysis. Until now, this challenge has been tackled from a batch learning perspective in which data is assumed to be at rest, and where models do not continuously integrate new information into already constructed models. We present an approach closer to the Big Data and Internet of Things paradigms in which data is arriving continuously and where models learn incrementally, achieving significant enhancements in terms of data processing (time, memory and computational costs), and obtaining competitive performances. This work compares and examines the hourly electrical power prediction of several streaming regressors, and discusses about the best technique in terms of time processing and performance to be applied on this streaming scenario.
\end{abstract}  


\end{frontmatter}


\section{Introduction}

\subsection{The Electrical Power Prediction for Combined Cycle Power Plants}

The efficiency in combined cycle power plants (CCPPs) is a key issue, as it is revealed in a recent report \cite{blackveatch2018} where shows that in the next decade, the number of projects involving combined cycle technology will increase by a $3.1\%$, and this estimation is based on the high efficiency of CCPPs. The electrical power prediction in CCPPs encompasses numerous factors that should be considered to achieve an accurate estimation. The operators of a power grid often predict the power demand based on historical data and environmental factors, such as temperature, pressure, and humidity. Then, they compare these predictions with available resources, such as coal, natural gas, nuclear, solar, wind, or hydro power plants. Power generation technologies (e.g. solar and wind) are highly dependent on environmental conditions, and all generation technologies are subject to planned and unplanned maintenance. Thus, the challenge for a power grid operator is how to handle a shortfall in available resources versus actual demand.  The power output of a peaker power plant varies depending on environmental conditions, so the business problem is predicting the power output of a peaker power plant as a function of the environmental conditions -- since this would enable the grid operator to make economic trade-offs about the number of peaker plants to turn on (or whether to buy expensive power from another grid).

The referred CCPP in this work uses two gas turbines (GT) and one steam turbine (ST) together to produce up to $50\%$ more electricity from the same fuel than a traditional simple-cycle plant. The waste heat from the GTs is routed to the nearby two STs, which generate extra power. In this real environment, a thermodynamical analysis compels thousands of nonlinear equations whose solution is near unfeasible, taking too many computational, memory and time costs. This barrier is overcome by using a machine learning based approach, which is a frequent alternative instead of thermodynamical approaches \cite{kesgin2005simulation}. Concretely, this work applies stream regression (SR) machine learning algorithms for a prediction analysis of a thermodynamic system, which is the mentioned CCPP. The correct prediction of its electrical power output is very relevant for the efficiency and economic operation of the plant, and maximizes the income from the available megawatt hours. The sustainability and reliability of the GTs depend highly on this electrical power output prediction, above all when it is subject to constraints of high profitability and contractual liabilities.

\subsection{Stream Learning in the Big Data Era}

The \textit{Big Data} paradigm has gained momentum last decade because of its promise to deliver valuable insights to many real-world applications \cite{zhou2014big}. With the advent of this emerging paradigm comes not only an increase in the volume of available data, but also the notion of its arrival velocity, that is, these real-world applications generate data in real-time at rates faster than those that can be handled by traditional systems. This situation leads us to assume that we have to deal with a potentially infinite and ever-growing dataset that may arrive continuously (\textit{stream learning}, SL) in batches of instances or instance by instance, in contrast to traditional systems (\textit{batch learning}) where there is free access to all historical data. These traditional processing systems assume that data is at rest and simultaneously accessed. For instance, database systems can store large collections of data and allow users to run queries or transactions. The models based on batch processing do not continuously integrate new information into already constructed models but instead regularly reconstruct new models from the scratch. However, the incremental learning that is carried out by SL presents advantages for this particular stream processing by continuously incorporating information into its models, and traditionally aim for minimal processing time and space. Because of its ability of continuous large-scale and real-time processing, incremental learning has recently gained more attention in the context of \textit{Big Data} \cite{chen2014big}. SL also presents many new challenges and poses stringent conditions \cite{domingos2003general}: only a single sample (or a small batch of instances) is provided to the learning algorithm at every time instant, a very limited processing time, a finite amount of memory, and the necessity of having trained models at every scan of the streams of data. In addition, these streams of data may evolve over time and may be occasionally affected by a change in their data distribution (\textit{concept drift})\cite{8496795}, forcing the system to learn under non-stationary conditions.

We can find many examples of real-world SL applications \cite{alippi2014intelligence}, such as mobile phones, industrial process controls, intelligent user interfaces, intrusion detection, spam detection, fraud detection, loan recommendation, monitoring and traffic management, among others \cite{vzliobaite2016overview}. In this context, the \textit{Internet of Things} (\textit{IoT}) has become one of the main applications of SL \cite{de2016iot}, since it is producing huge quantity of data continuously in real-time. The \textit{IoT} is defined as sensors and actuators connected by networks to computing systems \cite{McKinsey}, which monitors and manages the health and actions of connected objects or machines in real-time. Therefore, stream data analysis is becoming a standard to extract useful knowledge from what is happening at each moment, allowing people or organizations to react quickly when inconveniences emerge or when new trends appear, helping them to increase their performance. 

\subsection{CCPPs and Stream Learning Regression}

The task of power output prediction can be seen as a process based on data streams, as we will show in this work. Even though the work \cite{tufekci2014prediction} is perfectly adequate under specific conditions which allow a batch processing, and where the author assumed the possibility of storing all the historical data to process it and predict the electrical power output with machine learning regression algorithms, we tackle the same problem from a contemporary streaming perspective. 

In this work we have considered a CCPP as a practical case of \textit{IoT} application, where different sensors provide the required data to efficiently predict in real-time the full load electrical power output (see Figure \ref{CCPP_scheme}). In fact, all data generated by \textit{IoT} applications can be considered as streaming data since it is obtained in specific intervals of time. Power generation is a complex process, and understanding and predicting power output is an important element in managing a CCPP and its connection to the power grid.

Our view is closer to a reality where fast data can be huge, is in motion, and is closely connected, and where there are limited resources (e.g. time, memory) to process it. While it does not seem appropriate to retrain the learning algorithms every time new instances are available (what occurs in batch processing), a streaming perspective introduces significant enhancements in terms of data processing (less time and computational costs), algorithms training (they are updated every time new instances come), and presents a modernized vision of a CCPP considering it as an \textit{IoT} application, and as a part of the Industry $4.0$ paradigm \cite{lasi2014industry}. To the best of our knowledge, this is the first time that a SL approach is applied to CCPPs for electrical output prediction. This work could be widely replicated for other streaming prediction purposes in CCPPs, even more, it can serve as a practical example of SL application for modern electrical power industries that need to obtain benefits from the \textit{Big Data} and \textit{IoT} paradigms.

Our work uses some of the most known SR learning algorithms to successfully predict in an online manner the electrical power output by using a combination of input parameters defined by for GTs and STs (ambient temperature, vacuum, atmospheric pressure, and relative humidity). This work shows how the application of a SL perspective fits the purposes of a modern industry in which data flows constantly, analyzing the impact of several streaming factors (which should be considered before the streaming process starts) on the output prediction. It also compares the results represented by several error metrics and time processing of several SRs under different experiments, finding the most recommendable ones in the electrical power output prediction, aside from carrying out a statistical significance study. 

This work is organized as follows. Section \ref{related_work} provides a background about the topics of the manuscript. In Section \ref{mats_methods} materials and methods are presented, whereas Section \ref{comp_anal} describes the experimental work. Section \ref{discus} provides a discussion of the work, and then Section \ref{conc} finalizes by presenting the final conclusions of the work.  

\section{Related Work}\label{related_work}

The literature have undertaken related problems by using machine learning approaches. In \cite{kaya2012local,tufekci2014prediction} the authors successfully applied several regression methods to predict the full load electrical power output of a CCPP. A different approach for the same goal was investigated in \cite{rashid2015energy}, where the authors presented a novel approach using a particle swarm optimization \cite{kennedy2010particle} trained feedforward neural network to predict power plant output. In line with this last study, the work in \cite{manshad2016application} developed a new artificial neural network optimized by particle swarm optimization for dew point pressure prediction. In \cite{cavarzere2012application} the authors applied forecasting methodologies, including linear and nonlinear regression, to predict GT behavior over time, which allows planning maintenance actions and saving costs, and also because unexpected stops can be avoided. This work \cite{sekhon2008comparison} presents a comparison of two strategies for GT performance prediction, using statistical regression as technique to analyze dynamic plant signals. The prognostic approach to estimate the remaining useful life of GT engines before their next major overhaul was overcome in \cite{li2009gas}, where a combination of regression techniques were proposed to predict the remaining useful life of GT engines. In \cite{memon2015parametric} was showed that regression models were good estimators of the response variables to carry out parametric based thermo-environmental and exergoeconomic analyses of CCPPs. The same authors were involved in \cite{memon2014thermo} when using multiple polynomial regression models to correlate the response variables and predictor variables in a CCPP to carry out a thermo-environmental analysis. More recently, in \cite{tsoutsanis2017derivative} is presented a real-time derivative-driven regression method for estimating the performance of GTs under dynamic conditions. A scheme for performance-based prognostics of industrial GTs operating under dynamic conditions is proposed and developed in \cite{tsoutsanis2016dynamic}, where a regression method is implemented to locally represent the diagnostic information for subsequently forecasting the performance behavior of the engine.

Regarding the SL topic, many researches have focused on it due to its mentioned relevance, such as \cite{losing2018incremental,khamassi2018discussion,ramirez2017survey,gomes2017survey,tennant2017scalable}, and more recently in \cite{lobo2018dred,lobo2018evolving,almeida2018adapting,de2019overview}. The application of regression techniques to SL has been recently addressed in \cite{2018arXiv180205872B}, where the authors cover the most important online regression methods. The work \cite{krawczyk2017ensemble} deals with ensemble learning from data streams, and concretely it focused on regression ensembles. The authors of \cite{lughofer2017online} propose several criteria for efficient sample selection in case of SL regression problems within an online active learning context. In general, we can say that regression tasks in SL have not received as much attention as classification tasks, and this was spotlighted in \cite{ikonomovska2015online}, where researchers carried out an study and an empirical evaluation of a set of online algorithms for regression, which includes the baseline Hoeffding-based regression trees, online option trees, and an online least mean squares filter.

Next we present the materials and methods to carry out the experimental benchmark.

\section{Materials and Methods}\label{mats_methods}

\subsection{System Description}

\begin{figure}
	\centering
	\includegraphics[width=\columnwidth]{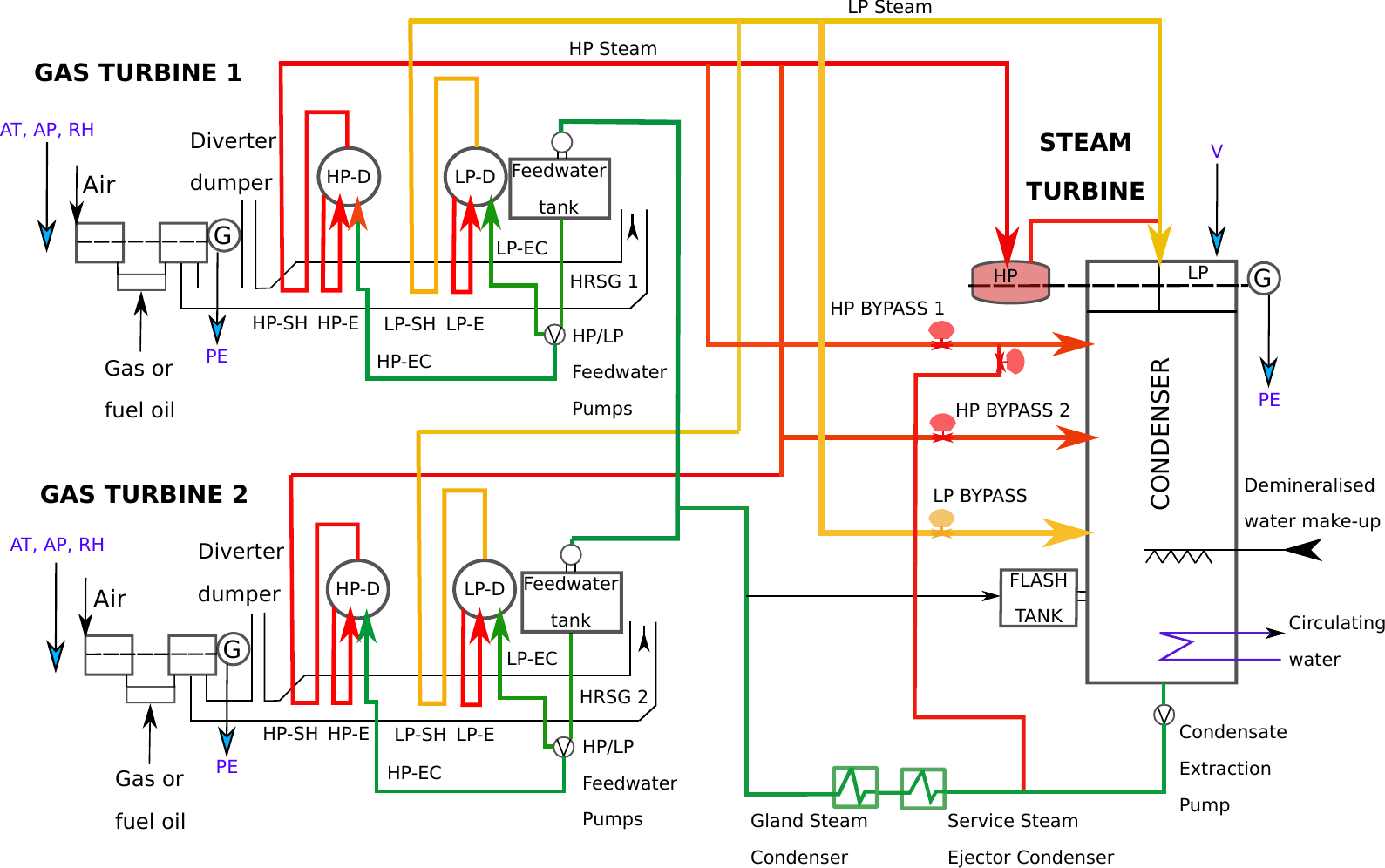}
	\caption{Layout of the combined cycle power plant based on \cite{tufekci2014prediction}. HP is High Pressure, LP is Low Pressure, D is Drum, G is Generator, SH is Super Heater, E is Evapo, EC is Eco, and HRSG is Heat Recovery Steam Generators. AT, AP, RH, V and PE are the variables described in Table \ref{variables}.}
	\label{CCPP_scheme}
\end{figure}

The proposed CCPP is composed of two GTs, one ST and two heat recovery steam generators. In a CCPP, the electricity is generated by GTs and STs, which are combined in one cycle, and is transferred from one turbine to another \cite{niu2008multivariable}. The CCPP captures waste heat from the GT to increase efficiency and the electrical output. Basically, how a CCPP works is as follows (see Figure \ref{CCPP_scheme}):
\paragraph{Gas turbine burns fuel}The GT compresses air and mixes it with fuel that is heated to a very high temperature. The hot air-fuel mixture moves through the GT blades, making them spin. The fast-spinning turbine drives a generator that converts a portion of the spinning energy into electricity
\paragraph{Heat recovery system captures exhaust}A Heat Recovery Steam Generator captures exhaust heat from the GT that would otherwise escape through the exhaust stack. The Heat Recovery Steam Generator creates steam from the GT exhaust heat and delivers it to the ST.
\paragraph{Steam turbine delivers additional electricity}The ST sends its energy to the generator drive shaft, where it is converted into additional electricity.

This type of CCPP is being installed in increasing number of plants around the world where there is access to substantial quantities of natural gas \cite{CCPP_rep}. As it was reported in \cite{tufekci2014prediction}, the proposed CCPP is designed with a nominal generating capacity of $480$ megawatts, made up of $2$ X $160$ megawatts ABB $13$E$2$ GTs, $2$ X dual pressure Heat Recovery Steam Generators and $1$ X $160$ megawatts ABB ST. GT load is sensitive to the ambient conditions; mainly ambient temperature (AT), atmospheric pressure (AP), and relative humidity (RH). However, ST load is sensitive to the exhaust steam pressure (or vacuum, V). These parameters of both GTs and STs are used as input variables, and the electrical power generating by both GTs and STs is used as a target variable in the dataset of this study. All of them are described in Table \ref{variables} and correspond to average hourly data received from the measurement points by the sensors denoted in Figure \ref{CCPP_scheme}.

\subsection{The Stream Learning Process}\label{SL_process}
 
We define a SL process as one that generates on a given stream of training data $s_{1},s_{2},s_{3},...,s_{t}$ a sequence of models $h_{1},h_{2},h_{3},...,h_{t}$. In our case $s_{i}$ is labeled training data $s_{i}=(x_{i},y_{i}) \in \R^{n} \times \{1,...,C\}$ and $h_{i} \colon \R^{n}\{1,...,C\}$ is a model function solely depending on $h_{i-1}$ and the recent $p$ instances $s_{i},...,s_{i-p}$ with $p$ being strictly limited (in this work $p=1$, representing a real case with a very stringent use case of online learning). The learning process in streaming is incremental \cite{losing2018incremental}, which means that we have to face the following challenges:
\begin{itemize}
	\item The stream algorithm adapts/learns gradually (i.e. $h_{i+1}$ is constructed based on $h_{i}$ without a complete retraining),
	\item Retains the previously acquired knowledge avoiding the effect of catastrophic forgetting \cite{chen2016lifelong}, and
	\item Only a limited number of $p$ training instances are allowed to be maintained. In this work we have applied a real SL approach under stringent conditions in which instance storing is not allowed.
\end{itemize}

\begin{figure}[h!]
	\centering
	\includegraphics[width=\columnwidth]{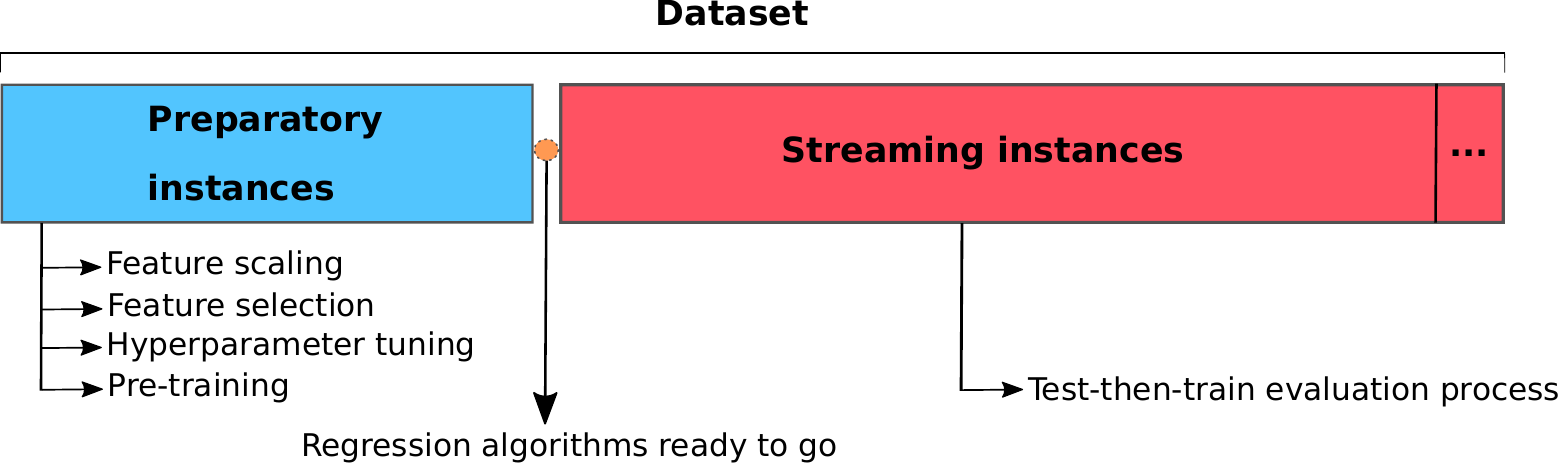}
	\caption{Scheme of the SL process of this work.}
	\label{stream_process}
\end{figure}

Therefore, data-intensive applications often work with transient data: some or all of the input instances are not available from memory. Instances in the stream arrive online (frequently one instance at a time) and can be read at most once, which constitutes the strongest constraint for processing data streams, and the system has to decide whether the current instance should be discarded or archived. Only selected past instances can be accessed by storing them in memory, which is typically small relative to the size of the data streams. When designing SL algorithms, we have to take several algorithmic and statistical considerations into account. For example, we have to face the fact that, as we cannot store all the inputs, we cannot unwind a decision made on past data. In batch learning processing, we have free access to all historical data gathered during the process, and then we can apply ``preparatory techniques'' such as pre-processing, feature selection or statistical analysis to the dataset, among others (see Figure \ref{stream_process}). Yet the problem with stream processing is that there is no access to the whole past dataset, and we have to opt for one of the following strategies. The first one is to carry out the preparatory techniques every time a new batch of instances or one instance is received, which increments the computational cost and time processing; it may occur that the process flow cannot be stopped to carry out this preparatory process because new instances continue arriving, which can be a challenging task. The second one is to store a first group of instances (preparatory instances) and carry out those preparatory techniques and data stream analysis, applying the conclusions to the incoming instances. This latter case is very common when streaming is applied to a real environment and it has been adopted by this work. We will show later how the selection of the size of this first group of instances (it might depend on the available memory or the time we can take to collect or process these data) can be crucial to achieve a competitive performance in the rest of the stream.

Once these first instances have been collected, in this work we will apply three common preparatory techniques before the streaming process starts in order to prepare our SRs:
\paragraph{Feature selection}It is one of the core concepts in machine learning that hugely impacts on the performance of models; irrelevant or partially relevant features can negatively impact model performance. Feature selection can be carried out automatically or manually, and selects those features which contribute most to the target variable. Its goal is to reduce overfitting, to improve the accuracy, and to reduce time training. In this work we will show how the feature selection impacts on the final results.
\paragraph{Hyper-parameter tuning}A hyper-parameter is a parameter whose value is set before the learning process begins, and this technique tries to choose a set of optimal hyper-parameters for a learning algorithm in order to prevent overfitting and to achieve the maximum performance. There are two main different methods for optimizing hyper-parameters: grid search and random search. The first one works by searching exhaustively through a specified subset of hyper-parameters, guaranteeing to find the optimal combination of parameters supplied, but the drawback is that it can be very time consuming and computationally expensive. The second one searches the specified subset of hyper-parameters randomly instead of exhaustively, being its major benefit that decreases processing time, but without guaranteeing to find the optimal combination of hyper-parameters. In this work we have opted for a random search strategy considering a real scenario where computational resources and time are limited.
\paragraph{Pre-training}Once we have isolated a set of instances to carry out the previous techniques, why do not we also use these instances to train our SRs before the streaming process starts? As we will see in Section \ref{stream_eval}, where the \textit{test-then-train} evaluation is explained, by carrying out a pre-training process our algorithms will obtain a better prediction than if they were tested after being trained by one single instance.

\subsection{Stream Regression Algorithms}
A SL algorithm, like every machine learning method, estimates an unknown dependency between the independent input variables, and a dependent target variable, from a dataset. In our work, SRs predict the electrical power output of a CCPP from a dataset which consists of couples $(\textbf{x}_{t},y_{t})$ (i.e. an instance), and they build a mapping function $\hat{y_{t}}=(\textbf{x}_{t},y_{t})$ by using these couples. Their goal is to select the best function that minimizes the error between the actual output $(y_{t})$ of a system and predicted output $(\hat{y_{t}})$ based on instances of the dataset (training instances).

The prediction of a real value (regression) is a very frequent problem researched in the machine learning field \cite{draper2014applied}, thus they are used to control response of a system for predicting a numeric target feature. Many real-world challenges are solved as regression problems, and evaluated using machine learning approaches to develop predictive models. Concretely, the following proposed algorithms have been specifically designed to run on real-time, being capable of learning incrementally every time a new instance arrives. They have been selected due to their wide use in the SL community, and because their implementation can be easily found in three well-known Python frameworks, scikit-multiflow \cite{montiel2018scikit}, scikit-garden\footnote{https://github.com/scikit-garden/scikit-garden} and scikit-learn \cite{pedregosa2011scikit}.

\paragraph{Passive-Aggressive Regressor (\texttt{PAR})}The Passive-Aggressive technique focuses on the target variable of linear regression functions, $\hat{y_{t}}=\textbf{w}_{t}^{T} \cdot \textbf{x}_{t}$, where $\textbf{w}_{t}$ is the incrementally learned vector. When a prediction is made, the algorithm receives the true target value $y_{t}$ and suffers an instantaneous loss ($\varepsilon$-insensitive hinge loss function). This loss function was specifically designed to work with stream data and it is analogous to a standard hinge loss. The role of $\varepsilon$ is to allow a low tolerance of prediction errors. Then, when a round finalizes, the algorithm uses $\textbf{w}_{t}$ and the instance $(\textbf{x}_{t},y_{t})$ to produce a new weight vector $\textbf{w}_{t+1}$, which will be used to extend the prediction on the next round. In \cite{crammer2006online} the adaptation to learn regression is explained in detail.

\paragraph{Stochastic Gradient Descent Regressor (\texttt{SGDR})}Linear model fitted by minimizing a regularized empirical loss with stochastic gradient descent (SGD) \cite{bottou2010large} is one of the most popular algorithms to perform optimization for machine learning methods. There are three variants of gradient descent: batch gradient descent (BGD), SGD, and mini-batch gradient descent (mbGD). They differ in how much data we use to compute the gradient of the objective function; depending on the amount of data, we make a trade-off between the accuracy of the parameter update and the time it takes to perform an update. BGD and mbGD perform redundant computations for large datasets, as they recompute gradients for similar instances before each parameter update. SGD does away with this redundancy by performing one update at a time; it is therefore usually much faster and it is often used to learn online \cite{zhang2004solving}.

\paragraph{Multi-layer Perceptron Regressor (\texttt{MLPR})}Multi-layer Perceptron (MLP) \cite{rumelhart1988learning} learns a non-linear function approximator for either classification or regression. MLPR uses a MLP that trains using backpropagation with no activation function in the output layer, which can also be seen as using the identity function as activation function. It uses the square error as the loss function, and the output is a set of real values.

\paragraph{Regression Hoeffding Tree (\texttt{RHT})}It is a regression tree that is able to perform regression tasks. A Hoeffding Tree (HT) or a Very Fast Decision Tree (VFDT) \cite{domingos2000mining} is an incremental anytime decision tree induction algorithm that is capable of learning from massive data streams, assuming that the distribution generating instances does not change over time, and exploiting the fact that a small instance can often be enough to choose an optimal splitting attribute. The idea is supported mathematically by the Hoeffding bound, which quantifies the number of instances needed to estimate some statistics within the goodness of an attribute. A RHT can be seen as a Hoeffding Tree with two modifications: instead of using information gain to split, it uses variance reduction; and instead of using majority class and naive bayes at the leaves, it uses target mean, and the perceptron \cite{ikonomovska2011learning}.

\paragraph{Regression Hoeffding Adaptive Tree (\texttt{RHAT})}In this case, \texttt{RHAT} is like \texttt{RHT} but using ADWIN \cite{bifet2007learning} to detect drifts and perceptron to make predictions. As it has been previously mentioned, streams of data may evolve over time and may show a change in their data distribution, what provokes that learning algorithms become obsolete. By detecting these drifts we are able to suitably update our algorithms to the new data distribution \cite{khamassi2018discussion}.

\paragraph{Mondrian Tree Regressor (\texttt{MTR})}The \texttt{MTR}, unlike standard decision tree implementations, does not limit itself to the leaf in making predictions. It takes into account the entire path from the root to the leaf and weighs it according to the distance from the bounding box in that node. This has some interesting properties such as falling back to the prior mean and variance for points far away from the training data. This algorithm has been adapted by the scikit-garden framework to serve as a regressor algorithm.

\paragraph{Mondrian Forest Regressor (\texttt{MFR})}A \texttt{MFR} \cite{lakshminarayanan2014mondrian} is an ensemble of \texttt{MTR}s. As in any ensemble of learners, the variance in predictions is reduced by averaging the predictions from all learners (Mondrian trees). Ensemble-based methods are among the most widely used techniques for data streaming, mainly due to their good performance in comparison to strong single learners while being relatively easy to deploy in real-world applications \cite{gomes2017survey}.

\section{Comparative Analysis}\label{comp_anal}

\subsection{Dataset Description and Exploratory Analysis}\label{data_desc}
The dataset contains $9,568$ data points collected from a CCPP over $6$ years ($2006-2011$), when the power plant was set to work with full load over $674$ different days. The \textit{Data and Source Code Availability} section at the end of the manuscript contains the details and references of the dataset. Features described in Table \ref{variables} consist of hourly average ambient temperature, ambient pressure, relative humidity, and exhaust vacuum to predict the net hourly electrical energy output of the plant. Although it is a problem that has already been successfully tackled from a batch processing perspective \cite{tufekci2014prediction} due to its manageable number of instances and its data arriving rate, it could be easily transposed to a streaming scenario in which the available data would be huge (instances collected over many years) and in which the data arriving rate would be very constrained (e.g. instances received every second). This more realistic \textit{IoT} scenario would allow CCPPs to manage a \textit{Big Data} approach, being able to predict the electrical output every time (e.g. every second) new data is available, and detecting anomalies long before in order to take immediate action.

\begin{table}[h!]
	\vspace{0.5cm}	
	\centering
	\resizebox{\textwidth}{!}{
		\begin{tabular}{@{}ccccc@{}}
			\toprule
			VARIABLES & ABBREVIATIONS & DESCRIPTIONS & RANGES & TYPES \\ 
			\midrule
			Ambient Temperature & AT & Measured in whole degrees in Celsius & $1.81-37.11$ & Input \\
			Atmospheric Pressure & AP & Measured in units of milibars & $992.89-1033.30$ & Input \\
			Relative Humidity & RH & Measured as a percentage & $25.56-100.16$ & Input \\
			Vacuum (Exhaust Steam Pressure) & V & Measured in cm Hg & $25.36-81.56$ & Input \\
			Full Load Electrical Power Output & PE & Measured in megawatts & $420.26-495.76$ & Target \\ \bottomrule
	\end{tabular}}
	\caption{Input and target variables of the dataset.}
	\label{variables}
\end{table}

As we can see in Table \ref{variables}, our dataset highly varies in magnitudes, units and ranges. Feature scaling can vary our results a lot while using certain algorithms and have a minimal or no effect in others. It is recommendable to scale the features when algorithms compute distances (very often Euclidean distances) or assume normality. In this work we have opted for the min-max scaling method which brings the value between $0$ and $1$.

\begin{figure}[h!]
	\centering
	\includegraphics[width=0.8\columnwidth]{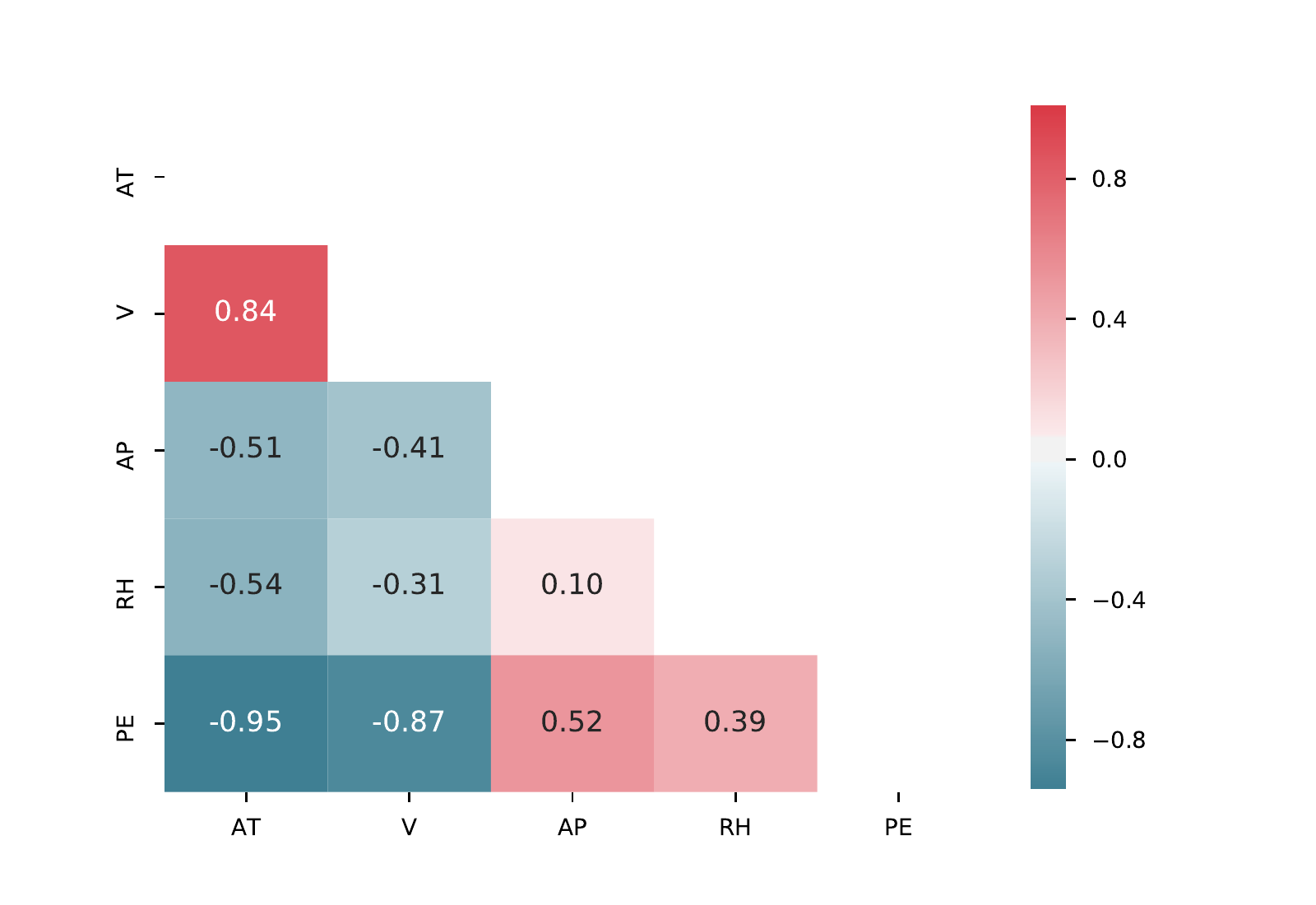}
	\caption{Heat map for visualizing the correlation between features.}
	\label{heat}
\end{figure}

\begin{figure}[h!]
	\centering
	\includegraphics[width=0.75\columnwidth]{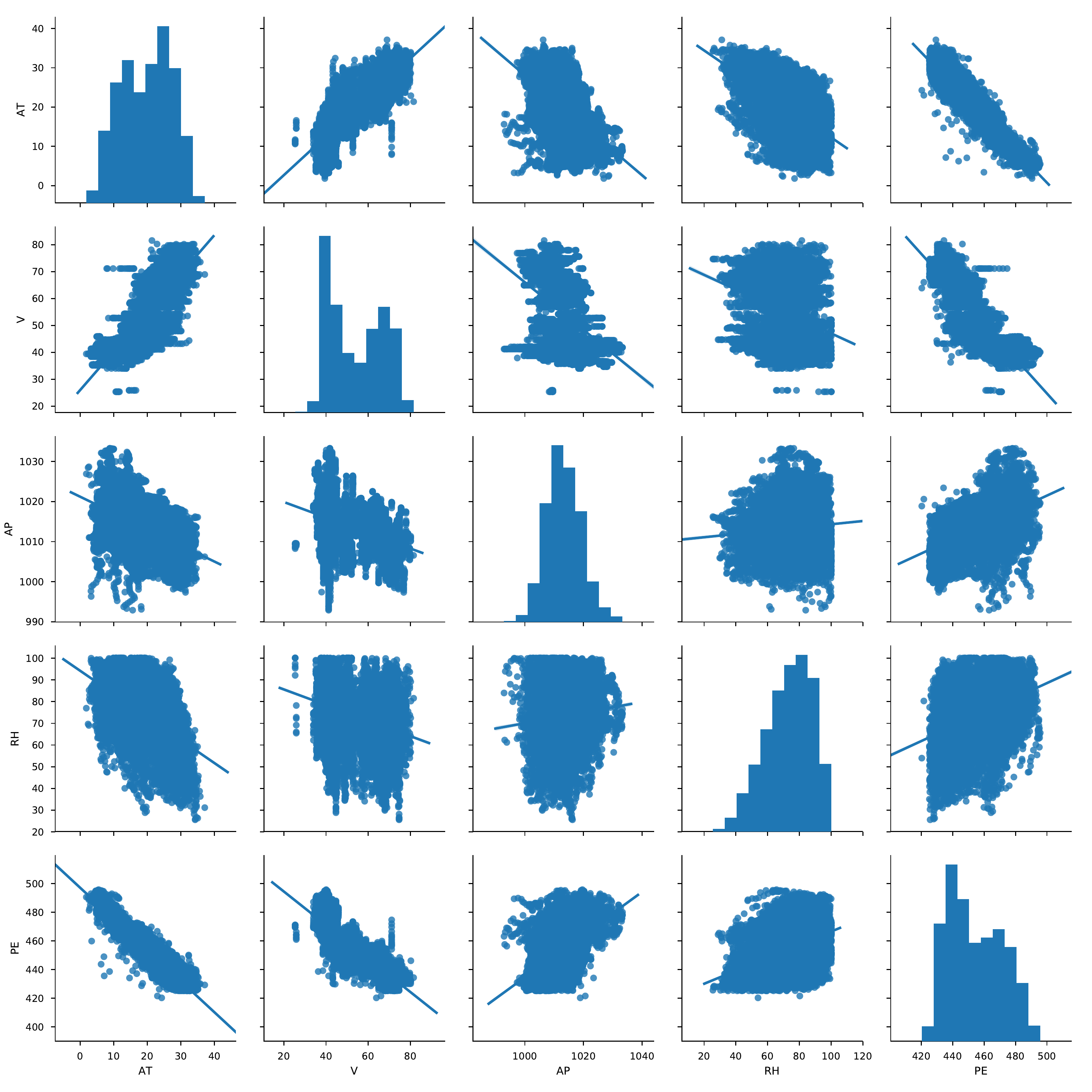}
	\caption{Scatter diagram for visualizing the correlation between features, and the linear regression model fit to the data.}
	\label{corr}
\end{figure}

The input variables (AT, V, AP, RH) affect differently the target variable (PE). Figure \ref{corr} shows the correlation between the input and the target variables. On the one hand, we observe how an increase in AT produces a decrease in PE, with a minimal vertical spread of scatter points which indicate a strong inverse relationship between them. This conclusion is supported by a correlation value of $-0.95$ in Figure \ref{heat}. In fact, there are some studies about GTs \cite{arrieta2005influence,de2011gas,erdem2006case} which show the effect of AT on the performance of CCPPs. The performance reduction due to an increase in temperature is known to stem from the decrease in the density of inlet air. 

On the other hand, we can see how with an increase in V produces a decrease in PE, and it can be also said that there is a strong inverse relationship between them. In this case, the spread is slightly larger than the variable AT, which hints at a slightly weaker relationship. This conclusion is also supported by a correlation value of $-0.87$ in Figure \ref{heat}. As it has been seen in Figure \ref{CCPP_scheme}, the CCPP uses a ST which leads to a considerable increase in total electrical efficiency. And when all other variables remain constant, V is known to have a negative impact on condensing-type turbine efficiency \cite{V_rep}. 

In the case of AP and RH, despite PE increases when they increase, Figure \ref{corr} depicts a big vertical spread of scatter points, which indicates weak positive relationships that are also confirmed in Figure \ref{heat}, where $0.52$ and $0.39$ respectively are shown as the correlation values for these variables. AP is also responsible for the density inlet air, and when all other variables remain constant PE increases with increasing AP \cite{arrieta2005influence}. In the case of RH, increases the exhaust-gas temperature of GTs which leads to an increase in the power generated by the ST \cite{arrieta2005influence,de2011gas,erdem2006case,lee2011development}.

\subsection{Prediction Metrics}\label{metrics}
The quality of a regression model is how well its predictions match up against actual values (target values), and we use error metrics to judge the quality of this model. They enable us to compare regressions against other regressions with different parameters. In this work we use several error metrics because each one gives us a complementary insight of the algorithms performance.

\paragraph{Mean Absolute Error (\texttt{MAE})}It is an easily interpretable error metric that does not indicate whether or not the model under or overshoots actual data. \texttt{MAE} is the average of the absolute difference between the predicted values and observed value. A small \texttt{MAE} suggests the model is great at prediction, while a large \texttt{MAE} suggests that the model may have trouble in certain areas. A \texttt{MAE} of $0$ means that the model is a perfect predictor of the outputs. \texttt{MAE} is defined as:
\begin{equation}\label{MAE}
MAE=\frac{1}{n}\sum_{j=1}^{n}\abs{y_{j}-\hat{y}_{j}}
\end{equation}

\paragraph{Root Mean Square Error (\texttt{RMSE})}It represents the sample standard deviation of the differences between predicted values and observed values (called residuals). \texttt{RMSE} is defined as:
\begin{equation}\label{RMSE}
RMSE=\sqrt{\frac{1}{n}\sum_{j=1}^{n}(y_{j}-\hat{y}_{j})^{2}}
\end{equation}
\texttt{MAE} is easy to understand and interpret because it directly takes the average of offsets, whereas \texttt{RMSE} penalizes the higher difference more than \texttt{MAE}. However, even after being more complex and biased towards higher deviation, \texttt{RMSE} is still the default metric of many models because loss function defined in terms of \texttt{RMSE} is smoothly differentiable and makes it easier to perform mathematical operations. Researchers will often use \texttt{RMSE} to convert the error metric back into similar units, making interpretation easier.

\paragraph{Mean Square Error (\texttt{MSE})}It is just like \texttt{MAE}, but squares the difference before summing them all instead of using the absolute value. We can see this difference in the equation below:
\begin{equation}\label{MSE}
MSE=\frac{1}{n}\sum_{j=1}^{n}(y_{j}-\hat{y}_{j})^{2}
\end{equation}
Because \texttt{MSE} is squaring the difference, will almost always be bigger than the \texttt{MAE}. Large differences between actual and predicted are punished more in \texttt{MSE} than in \texttt{MAE}. In case of outliers presence, the use of \texttt{MAE} is more recommendable since the outlier residuals will not contribute as much to the total error as \texttt{MSE}. 

\paragraph{R Squared (\texttt{$R^{2}$})}It is often used for explanatory purposes and explains how well the input variables explain the variability in the target variable. Mathematically, it is given by:
\begin{equation}\label{R2}
R^{2}=1-\dfrac{\sum_{j=1}^{n}(y_{j}-\hat{y}_{j})^{2}}{\sum_{j=1}^{n}(y_{j}-\bar{y}_{j})^{2}}
\end{equation}

\subsection{Streaming Evaluation Methodology}\label{stream_eval}

\begin{figure}[h!]
	\vspace{0.5cm}	
	\centering
	\includegraphics[width=\columnwidth]{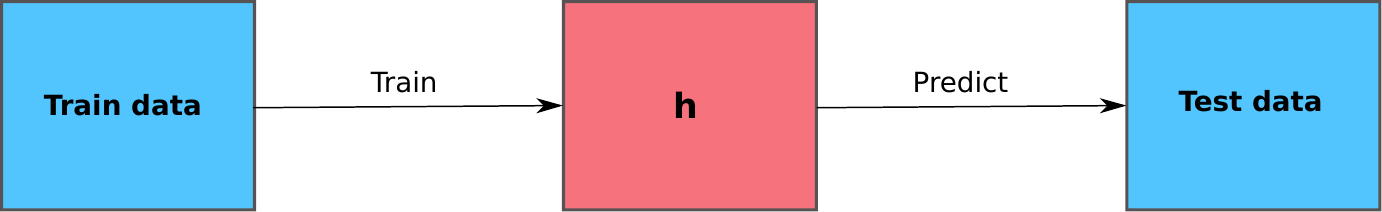}
	\caption{Traditional scheme when evaluating a non-incremental algorithm in batch processing mode.}
	\label{batch_processing}
\end{figure}

\begin{figure}[h!]
	\centering
	\includegraphics[width=\columnwidth]{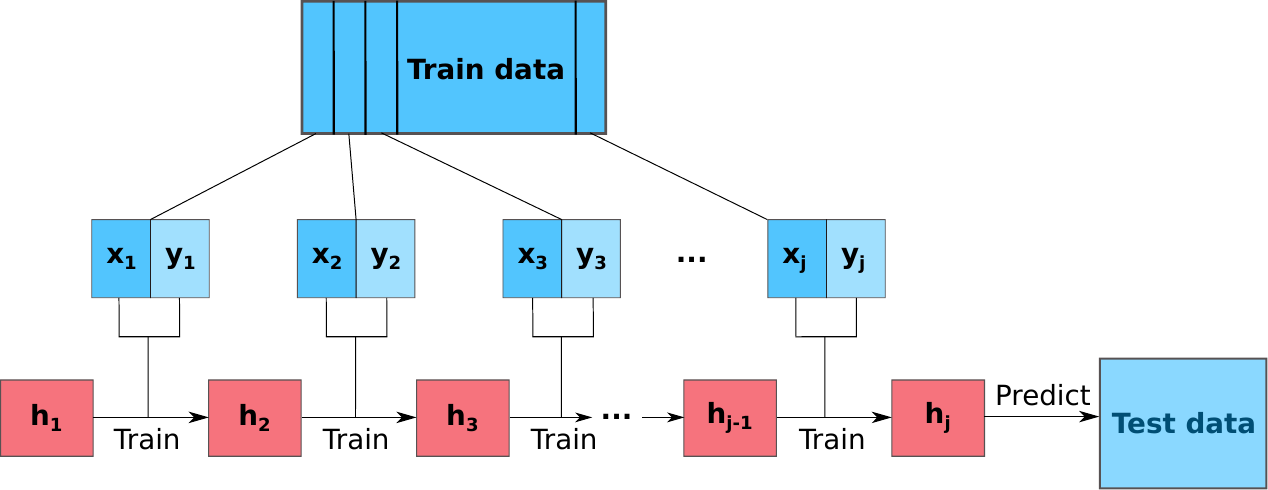}
	\caption{Training and testing of an incremental algorithm in batch processing mode. Note that only the last model is used for prediction.}
	\label{batch_incremental_processing}
\end{figure}

\begin{figure}[h!]
	\centering
	\includegraphics[width=\columnwidth]{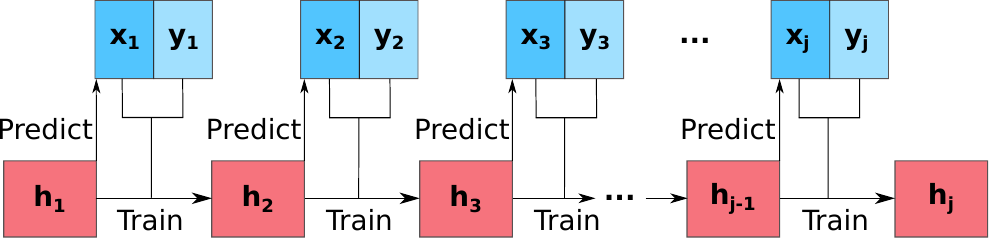}
	\caption{Stream learning (online learning) scheme with \textit{test-then-train} evaluation.}
	\label{stream_incremental_process}
\end{figure}

Evaluation is a fundamental task to know when an approach is outperforming another method only by chance, or when there is a statistical significance to that claim. In the case of SL, the methodology is very specific to consider the fact that not all data can be stored in memory (e.g. in online learning only one instance is processed at each time). Data stream regression is usually evaluated in the on-line setting, which is depicted in Figure \ref{stream_incremental_process}, and where data is not split into training and testing set. Instead, each model predicts subsequently one instance, which is afterwards used for the construction of the next model. In contrast, in the traditional evaluation for batch processing (see Figures \ref{batch_processing} and \ref{batch_incremental_processing} for non-incremental and incremental types respectively) all data used during training is obtained from the training set.

We have followed this evaluation methodology, proposed in \cite{gama2014survey,bifet2015efficient,MOA-Book-2018}, which recommends to follow these guidelines for streaming evaluation:

\paragraph{Error estimation}We have used an \textit{interleaved test-then-train} scheme, where each instance is firstly used for testing the model before it is used for training, and from this, the error metric is incrementally updated. The model is thus always being tested on instances it has not yet seen.

\paragraph{Performance evaluation measures}In Section \ref{metrics} we have already detailed the prediction metrics used in this work.

\paragraph{Statistical significance}When comparing regressors, it is necessary to distinguish whether a regressor is better than another one only by chance, or whether there is a statistical significance to ensure that. The analysis of variance (ANOVA test \cite{scheffe1999analysis}) is used to determine whether there are any statistically significant differences between the means of several independent groups. As in \cite{tufekci2014prediction}, in this work it is also used to compare results of machine learning experiments \cite{alpaydin2009introduction}. The idea is to test the null hypothesis (all regressors are equal), and the alternative hypothesis is that at least one pair is significantly different. In order to know how different one SR is from each other, we will also perform a multiple pairwise comparison analysis using Tukey's range test \cite{tukey1949comparing}.

\paragraph{Cost measure}We have opted for measuring the processing time (in seconds) of SRs in each experiment. The computer used in the experiments is based on a x86\_64 architecture with $8$ processors Intel(R) Core(TM) i7 at $2.70$GHz, and $32$ DDR$4$ memory running at $2,133$ MHz.

\subsection{Experiments}

We have designed an extensive experimental benchmark in order to find out the most suitable SR method for electrical power prediction in CCPPs, by comparing in terms of error metrics and time processing, $7$ widely used SRs. The \textit{Data and Source Code Availability} section at the end of the manuscript contains the access to the source code for this experimentation. We have also carried out an ANOVA test to know about the statistical significance of the experiments, and a Tukey's test to measure the differences between SR pair-wises. 

The experimental benchmark has been divided into four different experiments (see Table \ref{exps_tab}) which have considered two preparatory sizes and two feature selection options, and it is explained in Algorithm \ref{exp_process}. The idea is to observe the impact of the number of instances selected for the preparatory phase when the streaming process finalizes, and also to test the relevance of the feature selection process in this streaming scenario. Each experiment has been run $25$ times, and the experimental benchmark has followed the scheme depicted in Figure \ref{stream_process}.

\begin{table}[h!]
	\vspace{0.5cm}	
	\centering
	\resizebox{\columnwidth}{!}{	
	\begin{tabular}{cc|cc}
		\multicolumn{2}{c|}{\multirow{2}{*}{\textbf{}}} & \multicolumn{2}{c}{PREPARATORY SIZES (\% OF THE DATASET)} \\
		\multicolumn{2}{c|}{} & $5\%$ & $20\%$ \\ \hline
		\multirow{2}{*}{FEATURE SELECTION} & True & Exp1 & Exp3 \\
		& False & Exp2 & Exp4
	\end{tabular}}
	\caption{The experimental benchmark for the comparison of the SRs.}
	\label{exps_tab}
\end{table}

The experiments have been carried out under the \textit{scikit-multiflow} framework \cite{montiel2018scikit}, which has been implemented in Python language \cite{oliphant2007python} due to its current popularity in the machine learning community. Inspired by the most popular open source Java framework for data stream mining, Massive Online Analysis (MOA) \cite{MOA-Book-2018}, \textit{scikit-multiflow} includes a collection of widely used SRs (\texttt{RHT} and \texttt{RHAT} have been selected for this work), among other streaming algorithms (classification, clustering, outlier detection, concept drift detection and recommender systems), datasets, tools, and metrics for SL evaluation. It complements \textit{scikit-learn} \cite{pedregosa2011scikit}, whose primary focus is batch learning (despite the fact that it also provides researchers with some SL methods: \texttt{PAR}, \texttt{SGDR} and \texttt{MLPR} have been selected for this work) and expands the set of machine learning tools on this platform. The \textit{scikit-garden} framework in turn complements the experiments by proving the \texttt{MTR} and \texttt{MFR} SRs.

Regarding the feature selection process, in contrast to the study carried out in \cite{tufekci2014prediction} where different subsets of features were tested manually, we have opted for an automatic process. It is based on the feature importance, which stems from its Pearson correlation \cite{benesty2009pearson} with the target variable: if it is higher than a threshold ($0.65$), then it will be considered for the streaming process. As this is a streaming scenario, and thus we do not know the whole dataset beforehand, we have carried out the feature selection process only with the preparatory instances in each of the $25$ runs for Exp1 and Exp3 experiments. After that, we have assumed this selection of features for the rest of the streaming process. 

Finally, for the hyper-parameter tuning process, we have optimized the parameters by using a randomized and cross-validated search on hyper-parameters provided by \textit{scikit-learn}\footnote{\url{https://scikit-learn.org/stable/modules/generated/sklearn.model\_selection.RandomizedSearchCV.html}}. In contrast to other common option called cross-validated grid-search, not all parameter values are tried out, but rather a fixed number of parameter settings is sampled from the specified distributions. The result in parameter settings is quite similar in both cases, while the run time for randomized search is drastically lower. As in the previous case, we have also carried out the hyper-parameter tuning process only with the preparatory instances in each of the $25$ runs for all experiments. After that, we have assumed again the set of tuned parameters for the rest of the streaming process. 

\begin{algorithm}[h!]
	\DontPrintSemicolon
	\SetAlgoLined
	\SetKwInOut{Input}{Input}
	\SetKwInOut{Output}{Output}
	
	Scaling features process
	
	\For{every feature\_selection in [True, False]}{

		\For{every preparatory\_size in [$5\%$, $20\%$]}{
		
			\For{every run in $25$}{
				Create stream learners=[\texttt{PAR}, \texttt{SGDR}, \texttt{MLPR}, \texttt{RHT}, \texttt{RHAT}, \texttt{MFR}, \texttt{MTR}]
				
				Initialize performance metrics and time processing variables
				
				Create partitions for preparatory and test-then-train phases by using \textit{preparatory\_size}
				
				\If{feature\_selection=True}{
					
					Perform feature selection process
				}
			
				\For{every regressor }{
					
					Perform hyper-parameter tuning process
					
					Evaluation process for performance metrics and time processing
				}
				
			}
		
		}

	}

	Statistical tests
	
	Best stream learner selection	 
	
	\caption{Experimental benchmark structure.}
	\label{exp_process}
\end{algorithm}

\subsection{Results}

In this section we present the results of the SRs following the evaluation methodology presented in Section \ref{stream_eval}. Tables \ref{res_exp1}, \ref{res_exp2}, \ref{res_exp3}, \ref{res_exp4} show the error metrics and time processing of each SR for the experiments $1$,$2$,$3$ and $4$ respectively. Tables \ref{hyper1}, \ref{hyper2}, \ref{hyper3}, and \ref{hyper4} compiles the most suitable parameters (hyper-parameters) for SRs in each experiment. Finally, we present the results of the one-way ANOVA test for each experiment (with $25$ runs) and SR, which is based on the $R^{2}$ error metric. The null hypothesis states that the means of all SRs in the experiment are equal, and the alternative hypothesis is that at least one pair is significantly different. In the next section we will discuss about these results and their interpretations.

\begin{table}[h!]		
	\centering
	\resizebox{\columnwidth}{!}{
	\begin{tabular}{@{}cccccc@{}}
		\toprule
		\textbf{SR} & \textbf{MSE} & \textbf{RMSE} & \textbf{MAE} & \boldmath{$R^{2}$} & \textbf{TIME (s)} \\ \midrule
		\texttt{PAR} & 0.007$\pm$0.011 & 0.062$\pm$0.049 & 0.062$\pm$0.049 & 0.872$\pm$0.070 & 2.97$\pm$0.71 \\
		\texttt{SGDR} & 0.008$\pm$0.013 & 0.070$\pm$0.056 & 0.070$\pm$0.056 & 0.829$\pm$0.123 & 2.31$\pm$0.38 \\
		\texttt{MLPR} & 0.005$\pm$0.007 & 0.055$\pm$0.041 & 0.055$\pm$0.041 & 0.901$\pm$0.011 & 9.23$\pm$7.78 \\
		\texttt{RHT} & 0.004$\pm$0.006 & 0.052$\pm$0.039 & 0.052$\pm$0.039 & 0.900$\pm$0.024 & 2.55$\pm$0.55 \\
		\texttt{RHAT} & 0.005$\pm$0.007 & 0.054$\pm$0.040 & 0.054$\pm$0.040 & 0.893$\pm$0.024 & 3.87$\pm$0.75 \\
		\texttt{MFR} & 0.021$\pm$0.029 & 0.109$\pm$0.091 & 0.109$\pm$0.091 & 0.592$\pm$0.203 & 107.06$\pm$49.95 \\
		\texttt{MTR} & 0.019$\pm$0.027 & 0.104$\pm$0.086 & 0.104$\pm$0.086 & 0.629$\pm$0.186 & 1.344$\pm$0.18 \\ \bottomrule
	\end{tabular}}
	\caption{Results of the experiment 1: feature selection with a $5\%$ of preparatory instances. Note that \texttt{RMSE}=\texttt{MAE} because all differences are equal.}
	\label{res_exp1}			
\end{table}

\newpage

\begin{table}[h!]
	\vspace{-1.5cm}
	\centering
	\resizebox{\columnwidth}{!}{
	\begin{tabular}{@{}cccccc@{}}
		\toprule
		\textbf{SR} & \textbf{MSE} & \textbf{RMSE} & \textbf{MAE} & \boldmath{$R^{2}$} & \textbf{TIME (s)} \\ \midrule
		\texttt{PAR} & 0.006$\pm$0.009 & 0.057$\pm$0.044 & 0.057$\pm$0.044 & 0.885$\pm$0.013 & 3.29$\pm$1.50 \\
		\texttt{SGDR} & 0.008$\pm$0.012 & 0.069$\pm$0.053 & 0.069$\pm$0.053 & 0.821$\pm$0.119 & 2.48$\pm$0.90 \\
		\texttt{MLPR} & 0.005$\pm$0.007 & 0.055$\pm$0.041 & 0.055$\pm$0.041 & 0.897$\pm$0.020 & 14.76$\pm$12.51 \\
		\texttt{RHT} & 0.005$\pm$0.007 & 0.052$\pm$0.040 & 0.052$\pm$0.040 & 0.876$\pm$0.047 & 3.55$\pm$0.70 \\
		\texttt{RHAT} & 0.005$\pm$0.007 & 0.054$\pm$0.04 & 0.054$\pm$0.040 & 0.884$\pm$0.038 & 5.42$\pm$0.94 \\
		\texttt{MFR} & 0.004$\pm$0.007 & 0.042$\pm$0.039 & 0.042$\pm$0.039 & 0.922$\pm$0.041 & 125.18$\pm$60.71 \\
		\texttt{MTR} & 0.012$\pm$0.021 & 0.076$\pm$0.073 & 0.076$\pm$0.073 & 0.754$\pm$0.171 & 1.49$\pm$0.23 \\ \bottomrule
	\end{tabular}}
	\caption{Results of the experiment 2: no feature selection with a $5\%$ of preparatory instances. Note that \texttt{RMSE}=\texttt{MAE} because all differences are equal.}
	\label{res_exp2}			
\end{table}

\begin{table}[h!]	
	\vspace{-0.5cm}		
	\centering
	\resizebox{\columnwidth}{!}{	
	\begin{tabular}{@{}cccccc@{}}
		\toprule
		\textbf{SR} & \textbf{MSE} & \textbf{RMSE} & \textbf{MAE} & \boldmath{$R^{2}$} & \textbf{TIME (s)} \\ \midrule
		\texttt{PAR} & 0.005$\pm$0.007 & 0.055$\pm$0.041 & 0.055$\pm$0.041 & 0.904$\pm$0.013 & 2.40$\pm$0.90 \\
		\texttt{SGDR} & 0.005$\pm$0.007 & 0.056$\pm$0.041 & 0.056$\pm$0.041 & 0.901$\pm$0.021 & 1.86$\pm$0.58 \\
		\texttt{MLPR} & 0.004$\pm$0.007 & 0.052$\pm$0.039 & 0.052$\pm$0.039 & 0.912$\pm$0.016 & 9.12$\pm$8.83 \\
		\texttt{RHT} & 0.004$\pm$0.006 & 0.050$\pm$0.037 & 0.050$\pm$0.037 & 0.914$\pm$0.010 & 2.07$\pm$0.28 \\
		\texttt{RHAT} & 0.004$\pm$0.007 & 0.052$\pm$0.039 & 0.052$\pm$0.039 & 0.909$\pm$0.010 & 3.48$\pm$0.57 \\
		\texttt{MFR} & 0.022$\pm$0.030 & 0.113$\pm$0.092 & 0.113$\pm$0.092 & 0.570$\pm$0.205 & 94.98$\pm$42.94 \\
		\texttt{MTR} & 0.024$\pm$0.031 & 0.120$\pm$0.093 & 0.120$\pm$0.093 & 0.539$\pm$0.178 & 1.11$\pm$0.16 \\ \bottomrule
	\end{tabular}}
	\caption{Results of the experiment 3: feature selection with a $20\%$ of preparatory instances. Note that \texttt{RMSE}=\texttt{MAE} because all differences are equal.}
	\label{res_exp3}		
	\vspace{0.25cm}								
\end{table}

\begin{table}[h!]	
	\vspace{-0.5cm}				
	\centering
	\resizebox{\columnwidth}{!}{
	\begin{tabular}{@{}cccccc@{}}
		\toprule
		\textbf{SR} & \textbf{MSE} & \textbf{RMSE} & \textbf{MAE} & \boldmath{$R^{2}$} & \textbf{TIME (s)} \\ \midrule
		\texttt{PAR} & 0.006$\pm$0.010 & 0.057$\pm$0.044 & 0.057$\pm$0.044 & 0.890$\pm$0.010 & 2.38$\pm$0.79 \\
		\texttt{SGDR} & 0.005$\pm$0.007 & 0.055$\pm$0.040 & 0.055$\pm$0.040 & 0.901$\pm$0.014 & 1.79$\pm$0.48 \\
		\texttt{MLPR} & 0.004$\pm$0.006 & 0.051$\pm$0.037 & 0.051$\pm$0.037 & 0.917$\pm$0.011 & 7.96$\pm$7.12 \\
		\texttt{RHT} & 0.004$\pm$0.006 & 0.048$\pm$0.036 & 0.048$\pm$0.036 & 0.917$\pm$0.023 & 3.12$\pm$0.43 \\
		\texttt{RHAT} & 0.005$\pm$0.007 & 0.055$\pm$0.041 & 0.055$\pm$0.041 & 0.892$\pm$0.039 & 5.16$\pm$1.18 \\
		\texttt{MFR} & 0.003$\pm$0.006 & 0.036$\pm$0.035 & 0.036$\pm$0.035 & 0.940$\pm$0.053 & 109.65$\pm$42.04 \\
		\texttt{MTR} & 0.011$\pm$0.019 & 0.075$\pm$0.070 & 0.075$\pm$0.070 & 0.776$\pm$0.126 & 1.21$\pm$0.21 \\ \bottomrule
	\end{tabular}}
	\caption{Results of the experiment 4: no feature selection with a $20\%$ of preparatory instances. Note that \texttt{RMSE}=\texttt{MAE} because all differences are equal.}
	\label{res_exp4}
\end{table}

Next, we will firstly provide the results of the hyper-parameter tuning process in each experiment. Tables \ref{hyper1}, \ref{hyper2}, \ref{hyper3}, and \ref{hyper4} show the values for the most relevant parameters of each SR. In case of any need for checking the details of any parameter, please refer to the frameworks described previously: \textit{scikit-learn} (\texttt{PAR}, \texttt{SGDR}, and \texttt{MLPR}), \textit{scikit-multiflow} (\texttt{RHT} and \texttt{RHAT}), and \textit{scikit-garden} (\texttt{MTR} and \texttt{MFR}). And secondly, we will show in Table \ref{feat_sel} the results of the feature selection process in experiments $1$ and $3$.

\begin{table}[h!]
	\vspace{1.0cm}	
	\centering
	\resizebox{0.7\columnwidth}{!}{%
		\begin{tabular}{@{}ccc@{}}
			\toprule
			\textbf{SR} & \textbf{PARAMETERS} & \textbf{VALUES} \\ \midrule
			\texttt{PAR} & C & 0.05 \\ \bottomrule
			\multirow{4}{*}{\texttt{SGDR}} & alpha & 0.1/0.01 \\
			& loss & \textit{epsilon\_insensitive} \\
			& penalty & \textit{L1/L2} \\
			& learning\_rate & \textit{constant/optimal} \\ \bottomrule
			\multirow{6}{*}{\texttt{MLPR}} & hidden\_layer\_sizes & (50,50) \\
			& activation & \textit{relu} \\
			& solver & \textit{adam/sgd} \\
			& learning\_rate & \textit{constant, invscaling, adaptive} \\
			& learning\_rate\_init & 0.005/0.001 \\
			& alpha & 0.000001-0.000000001 \\ \bottomrule
			\multirow{4}{*}{\texttt{RHT}} & grace\_period & 200 \\
			& split\_confidence & 0.0000001 \\
			& tie\_threshold & 0.05 \\
			& leaf\_prediction & \textit{perceptron} \\ \bottomrule
			\multirow{5}{*}{\texttt{RHAT}} & grace\_period & 200 \\
			& split\_confidence & 0.0000001 \\
			& tie\_threshold & 0.05 \\
			& leaf\_prediction & \textit{perceptron} \\
			& delta (ADWIN detector) & 0.002 \\ \bottomrule
			\multirow{2}{*}{\texttt{MTR}} & max\_depth & 10-100 \\
			& min\_samples\_split & 10 \\ \bottomrule
			\multirow{3}{*}{\texttt{MFR}} & max\_depth & 10-80 \\
			& min\_samples\_split & 10 \\
			& n\_estimators & 50/100 \\ \bottomrule
		\end{tabular}%
	}
	\caption{Hyper-parameter tuning results for SRs in the experiment 1.}
	\label{hyper1}
	\vspace{1.5cm}	
\end{table}

\begin{table}[h!]
	\vspace{1.0cm}	
	\centering
	\resizebox{0.7\columnwidth}{!}{%
		\begin{tabular}{@{}ccc@{}}
			\toprule
			\textbf{SR} & \textbf{PARAMETERS} & \textbf{VALUES} \\ \midrule
			\texttt{PAR} & C & 0.5/1.0 \\ \bottomrule
			\multirow{4}{*}{\texttt{SGDR}} & alpha & 0.00001-0.1 \\
			& loss & \textit{epsilon\_insensitive} \\
			& penalty & \textit{L1/L2} \\
			& learning\_rate & \textit{constant/optimal/invscaling} \\ \bottomrule
			\multirow{6}{*}{\texttt{MLPR}} & hidden\_layer\_sizes & (50,50)/(100,100) \\
			& activation & \textit{relu/tanh/identity} \\
			& solver & \textit{adam/sgd} \\
			& learning\_rate & \textit{constant, invscaling, adaptive} \\
			& learning\_rate\_init & 0.0005-0.05 \\
			& alpha & 0.00001-0.000000001 \\ \bottomrule
			\multirow{4}{*}{\texttt{RHT}} & grace\_period & 200 \\
			& split\_confidence & 0.0000001 \\
			& tie\_threshold & 0.05 \\
			& leaf\_prediction & \textit{perceptron} \\ \bottomrule
			\multirow{5}{*}{\texttt{RHAT}} & grace\_period & 200 \\
			& split\_confidence & 0.0000001 \\
			& tie\_threshold & 0.05 \\
			& leaf\_prediction & \textit{perceptron} \\
			& delta (ADWIN detector) & 0.002 \\ \bottomrule
			\multirow{2}{*}{\texttt{MTR}} & max\_depth & 20-90 \\
			& min\_samples\_split & 2/5/10 \\ \bottomrule
			\multirow{3}{*}{\texttt{MFR}} & max\_depth & 20-90 \\
			& min\_samples\_split & 2/5 \\
			& n\_estimators & 50/100 \\ \bottomrule
		\end{tabular}%
	}
	\caption{Hyper-parameter tuning results for SRs in the experiment 2.}
	\label{hyper2}
	\vspace{1.5cm}	
\end{table}

\begin{table}[h!]
	\vspace{1.0cm}	
	\centering
	\resizebox{0.7\columnwidth}{!}{%
		\begin{tabular}{@{}ccc@{}}
			\toprule
			\textbf{SR} & \textbf{PARAMETERS} & \textbf{VALUES} \\ \midrule
			\texttt{PAR} & C & 0.01 \\ \bottomrule
			\multirow{4}{*}{\texttt{SGDR}} & alpha & $0.001$ \\
			& loss & \textit{epsilon\_insensitive} \\
			& penalty & \textit{elasticnet/L1} \\
			& learning\_rate & \textit{constant} \\ \bottomrule
			\multirow{6}{*}{\texttt{MLPR}} & hidden\_layer\_sizes & $(50)/(100)$ \\
			& activation & \textit{relu} \\
			& solver & \textit{adam/sgd} \\
			& learning\_rate & \textit{constant, invscaling, adaptive} \\
			& learning\_rate\_init & $0.005$ \\
			& alpha & $0.00001,0.000001$ \\ \bottomrule
			\multirow{4}{*}{\texttt{RHT}} & grace\_period & $200$ \\
			& split\_confidence & $0.0000001$ \\
			& tie\_threshold & $0.05$ \\
			& leaf\_prediction & \textit{perceptron} \\ \bottomrule
			\multirow{5}{*}{\texttt{RHAT}} & grace\_period & $200$ \\
			& split\_confidence & $0.0000001$ \\
			& tie\_threshold & $0.05$ \\
			& leaf\_prediction & \textit{perceptron} \\
			& delta (ADWIN detector) & $0.002$ \\ \bottomrule
			\multirow{2}{*}{\texttt{MTR}} & max\_depth & $20-60$ \\
			& min\_samples\_split & $10$ \\ \bottomrule
			\multirow{3}{*}{\texttt{MFR}} & max\_depth & $20-60$ \\
			& min\_samples\_split & $10$ \\
			& n\_estimators & $50/100$ \\ \bottomrule
		\end{tabular}%
	}
	\caption{Hyper-parameter tuning results for SRs in the experiment 3.}
	\label{hyper3}
	\vspace{1.5cm}	
\end{table}

\begin{table}[h!]
	\vspace{1.0cm}	
	\centering
	\resizebox{0.7\columnwidth}{!}{%
		\begin{tabular}{@{}ccc@{}}
			\toprule
			\textbf{SR} & \textbf{PARAMETERS} & \textbf{VALUES} \\ \midrule
			\texttt{PAR} & C & $0.5/1.0$ \\ \bottomrule
			\multirow{4}{*}{\texttt{SGDR}} & alpha & $0.001-0.01$ \\
			& loss & \textit{epsilon\_insensitive} \\
			& penalty & \textit{L1/L2} \\
			& learning\_rate & \textit{constant/optimal} \\ \bottomrule
			\multirow{6}{*}{\texttt{MLPR}} & hidden\_layer\_sizes & $(100)/(500)/(50,50)/(100,100)$ \\
			& activation & \textit{relu/tanh} \\
			& solver & \textit{adam/sgd} \\
			& learning\_rate & \textit{constant, invscaling, adaptive} \\
			& learning\_rate\_init & $0.0005-0.05$ \\
			& alpha & $0.001-0.000000001$ \\ \bottomrule
			\multirow{4}{*}{\texttt{RHT}} & grace\_period & $200$ \\
			& split\_confidence & $0.0000001$ \\
			& tie\_threshold & $0.05$ \\
			& leaf\_prediction & \textit{perceptron} \\ \bottomrule
			\multirow{5}{*}{\texttt{RHAT}} & grace\_period & $200$ \\
			& split\_confidence & $0.0000001$ \\
			& tie\_threshold & $0.05$ \\
			& leaf\_prediction & \textit{perceptron} \\
			& delta (ADWIN detector) & 0.002 \\ \bottomrule
			\multirow{2}{*}{\texttt{MTR}} & max\_depth & $40-100$ \\
			& min\_samples\_split & $5$ \\ \bottomrule
			\multirow{3}{*}{\texttt{MFR}} & max\_depth & $30-100$ \\
			& min\_samples\_split & $5$ \\
			& n\_estimators & $50/100$ \\ \bottomrule 
		\end{tabular}%
	}
	\caption{Hyper-parameter tuning results for SRs in the experiment 4.}
	\label{hyper4}
	\vspace{1.5cm}	
\end{table}

\begin{table}[h!]
	\centering
	\resizebox{0.55\columnwidth}{!}{	
	\begin{tabular}{cc|cccc}
		\multicolumn{2}{c|}{\multirow{2}{*}{\textbf{}}} & \multicolumn{4}{c}{FEATURES} \\
		\multicolumn{2}{c|}{} & AT & AP & RH & V \\ \hline
		\multirow{2}{*}{EXPERIMENTS} & 1 & y & n & n & y \\
		& 2 & y & n & n & y
	\end{tabular}}
	\caption{Feature selection results in each experiment. Those selected features are represented with \textit{y} (yes), the rest with \textit{n} (no).}
	\label{feat_sel}
\end{table}

\begin{figure}[h!]
	\centering
	\subfigure[Exp1]{\includegraphics[width=0.49\columnwidth]{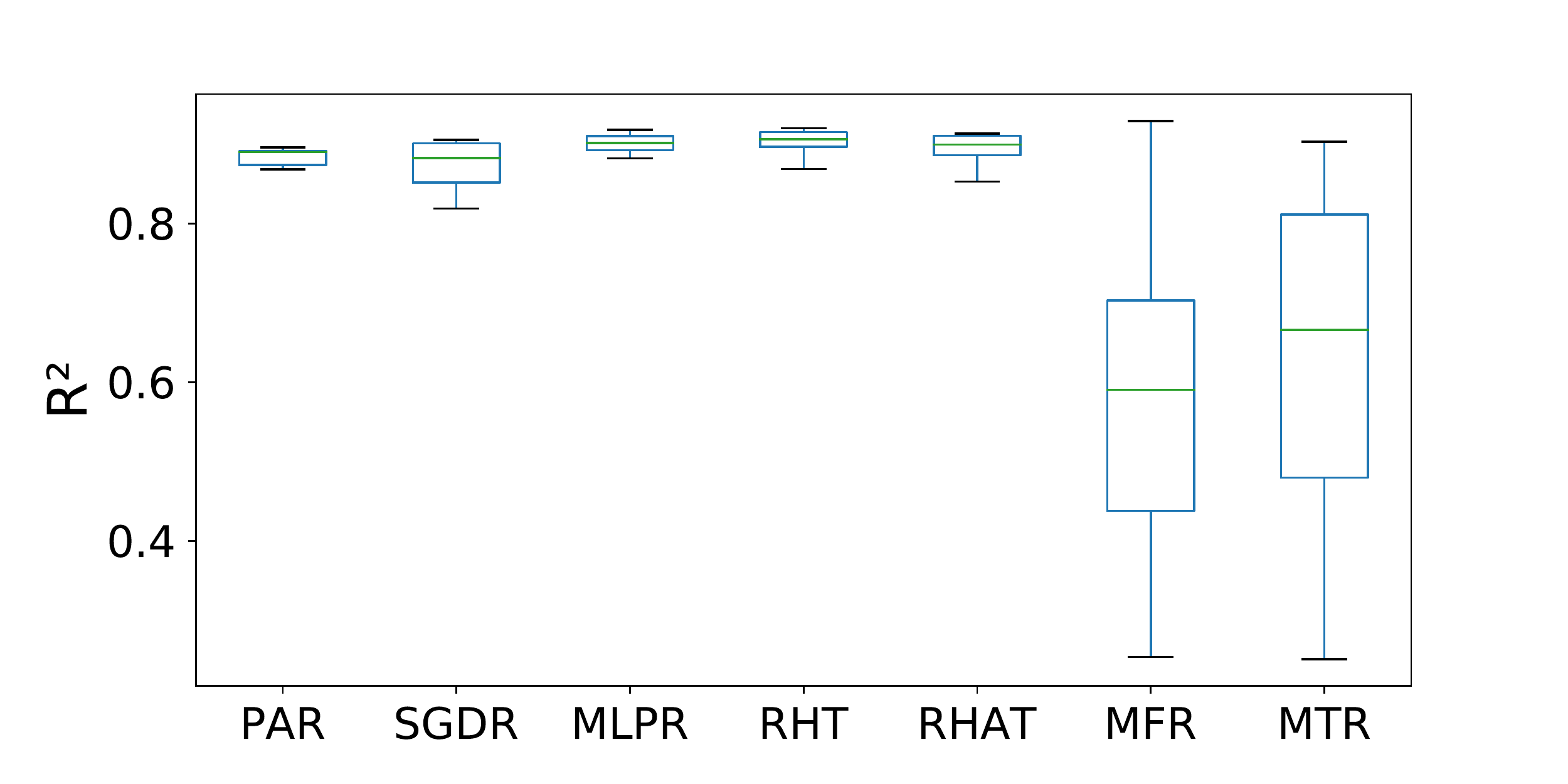}}
	\subfigure[Exp2]{\includegraphics[width=0.49\columnwidth]{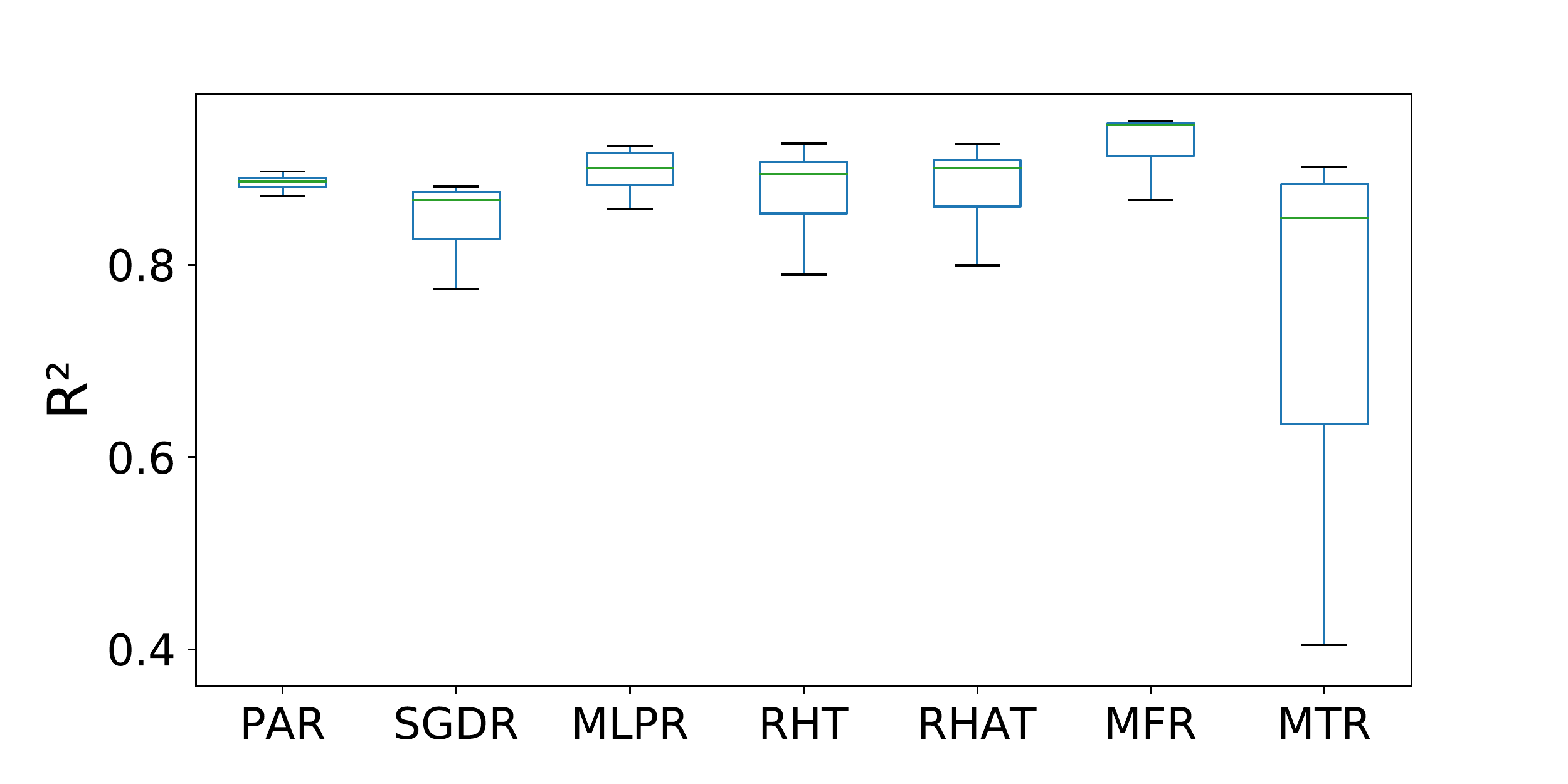}}
	\subfigure[Exp3]{\includegraphics[width=0.49\columnwidth]{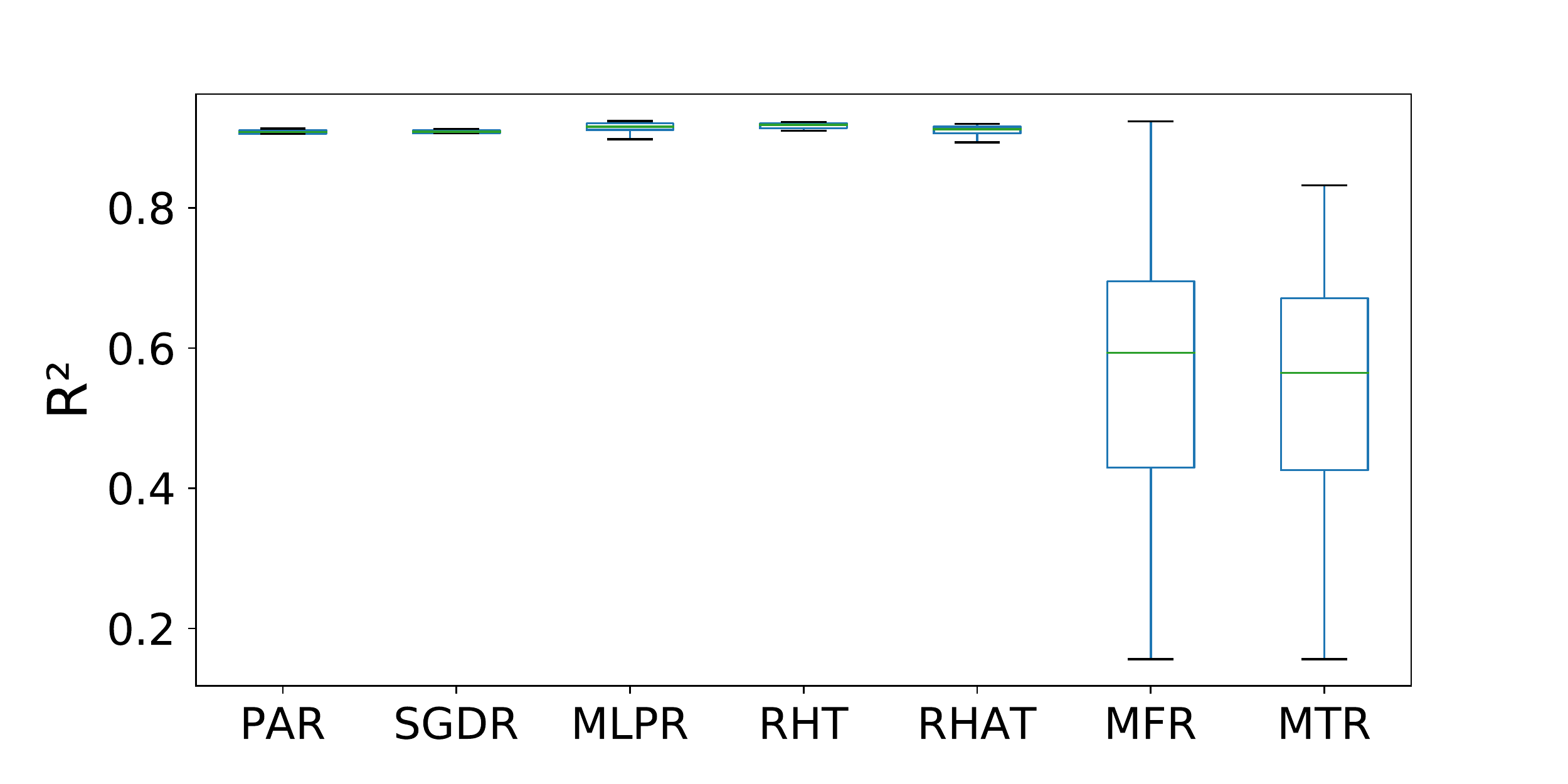}}
	\subfigure[Exp4]{\includegraphics[width=0.49\columnwidth]{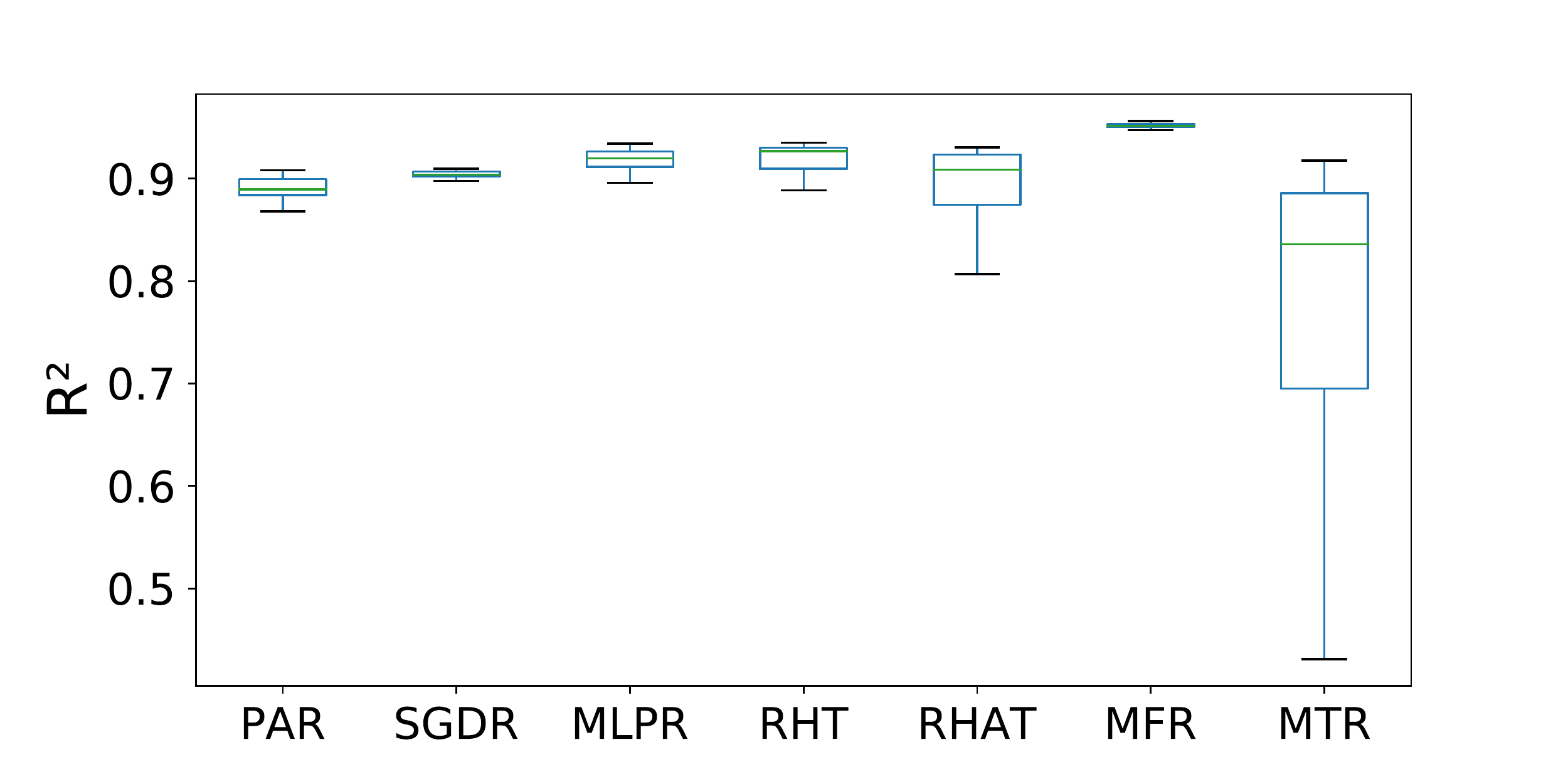}}	
	\caption{Boxplots of ANOVA tests for $R^{2}$ in each experiment.} 
	\label{anovas}
\end{figure}

Finally, Figure \ref{anovas} depicts the data distribution of ANOVA tests for each experiment. The p-values obtained from ANOVA analysis for each experiment ($p_{1}=5.82 \times e^{-25}$, $p_{2}=1.78 \times e^{-10}$, $p_{3}=2.46 \times e^{-42}$, $p_{4}=1.06 \times e^{-17}$) are significant ($p_{i}<0.05$), and therefore, we can conclude that there are significant differences among SRs performances. In order to know how different one SR is from each other, we have performed a Tukey's range test for each experiment (see \ref{tukeys}); results suggest that except some cases, all other pairwise comparisons reject null hypothesis and indicate statistical significant differences.

\section{Discussion}\label{discus}

We start the discussion by highlighting the relevance of having a representative set of preparatory instances in a SL process. As it was introduced in Section \ref{SL_process}, in streaming scenarios it is not possible to access all historical data, and then it is required to apply some strategy to make assumptions for the incoming data, unless a drift occurs (in which case it would be necessary an adaptation to the new distribution). One of these strategies consists of storing the first instances of the stream (preparatory instances) to carry out a set of preparatory techniques that make the streaming algorithms ready for the streaming process. We have opted for this strategy in our work, and in this section we will explain the impact of these preparatory process on the final performance of the SRs.

Preparatory techniques contribute to improve the performance of the SRs. Theoretically, by selecting a subset of features (feature selection) that contributes most to the prediction variable, we avoid irrelevant or partially relevant features that can negatively impact on the model performance. By selecting the most suitable parameters of algorithms (hyper-parameter tuning), we obtain SRs better adjusted to data. And by training our SRs before the streaming process starts (pre-training), we obtain algorithms ready for the streaming process with better performances. The drawback lies in the fact that as many instances we collect at the beginning of the process, as much time the preparatory techniques will need to be carried out. This is a trade-off that we should have to consider in each scenario, apart from the limits previously mentioned.

Regarding the number of the preparatory instances, as it often occurs with machine learning techniques, the more instances for training (or other purposes) are available, the better the performance of the SRs can be, because data distribution is better represented with more data and the SRs are more trained and adjusted to the data distribution. But on the other hand, the scenario usually poses limits in terms of memory size, computational capacity, or the moment in which the streaming process has to start, among others. Comparing the experiments $1$ and $3$ (see Tables \ref{res_exp1} and \ref{res_exp3} where the feature selection process was carried out and the preparatory instances were a $5\%$ and $20\%$ of the dataset respectively) with the experiments $2$ and $4$ (see Tables \ref{res_exp2} and \ref{res_exp4} where the feature selection process was not carried out and the preparatory instances were also a $5\%$ and $20\%$ of the dataset respectively), we observe how in almost all cases (except for \texttt{MTR} and \texttt{MFR} when feature selection was carried out) the error metrics improve when the number of preparatory instances is larger. Therefore, by setting aside a group of instances for preparatory purposes, we can generally achieve better results for these stream learners. 

In the case of the feature selection process, we deduce from the comparison between Tables \ref{res_exp1} and \ref{res_exp2} that this preparatory technique improves the performance of \texttt{RHT} and \texttt{RHAT}, and it also reduces their processing time. For \texttt{PAR}, \texttt{SGDR}, and \texttt{MLPR}, it achieves a similar performance but also reduces their processing time. Thus it is recommendable for all of them, except for \texttt{MTR} and \texttt{MFR}, when the preparatory size is $5\%$. In the case of the comparison between Tables \ref{res_exp3} and \ref{res_exp4}, this preparatory technique improves the performances of \texttt{PAR} and \texttt{RHAT}, and it also reduces o maintains their processing time. For \texttt{SGDR}, \texttt{MLPR} and \texttt{RHT} the performances and the processing times are very similar. Thus it is also recommendable for all of them, except again for \texttt{MTR} and \texttt{MFR}, when the preparatory size is $20\%$. In what refers to which features have been selected for the streaming process in the experiments $1$ and $3$, we see in Table \ref{feat_sel} how AT and V have been preferred over the rest by the hyper-parameter tuning method, which has also been confirmed in Section \ref{data_desc} due to their correlation with the target variable (PE). 

Regarding the selection of the best SR, Tables \ref{res_exp1}, \ref{res_exp2}, \ref{res_exp3}, and \ref{res_exp4} show how \texttt{MLP} and \texttt{RHT} show the best error metrics for both preparatory sizes when the feature selection process is carried out. When there is no a feature selection process, then the best error metrics are achieved by \texttt{MFR}. However, in terms of processing time, \texttt{SGDR} and \texttt{MTR} are the fastest stream learners. Due to the fact that we have to find a balance between error metric results and time processing, we recommend \texttt{RHT}. 

Finally, the ANOVA and Tukey's range tests have confirmed the statistical significance of this study (see Appendix \ref{tukeys}).

\section{Conclusion}\label{conc}

This work has presented a comparison of streaming regressors for electrical power prediction in a combined cycle power plant. This prediction problem had been tackled with the traditional thermodynamical analysis, which had shown to be computational and time processing expensive. However, some studies have addressed this problem by applying machine learning techniques, such as regression algorithms, and managing to reduce the computational and time costs. These new approaches have considered the problem under a batch learning perspective in which data is assumed to be at rest, and where regression models do not continuously integrate new information into already constructed models. Our work presents a new approach for this scenario in which data is arriving continuously and where regression models have to learn incrementally. This approach is closer to the emerging Big Data and IoT paradigms. 

The results show how competitive error metrics and processing times have been achieved when applying a SL approach to this specific scenario. Concretely, this work has identified \texttt{RHT} as the most recommendable technique to achieve the electrical power prediction. We have also highlighted the relevance of the preparatory techniques to make the streaming algorithms ready for the streaming process, and at the same time the importance of selecting properly the number of preparatory instances. Regarding the importance of the features, as in previous cases which tackled the same problem from a batch learning perspective, we do recommend to carry out a feature selection process for all SRs (except for \texttt{MTR} and \texttt{MFR}) because it reduces the streaming processing time and at the same time it is worthy due to the performance gain. Finally, as future work, we would like to transfer this SL approach to other processes in combined cycle power plants, and even to other kinds of electrical power plants.

\section*{Acknowledgements} \label{acknow}

This work was supported by the EU project \textit{iDev40}. This project has received funding from the ECSEL Joint Undertaking (JU) under grant agreement No 783163. The JU receives support from the European Union's Horizon 2020 research and innovation programme and Austria, Germany, Belgium, Italy, Spain, Romania. It has also been supported by the Basque Government (Spain) through the project \textit{VIRTUAL} (KK-2018/00096).

\section*{Data and Source Code Availability}\label{data_aval}
Source code and dataset related to this article can be found at: \newline https://github.com/TxusLopez/Streaming\_CCPP. The CCPP dataset has been taken from https://github.com/YungChunLu/UCI-Power-Plant, and originally was used in \cite{tufekci2014prediction} and taken from the UCI repository at:\newline https://archive.ics.uci.edu/ml/datasets/combined+cycle+power+plant.

\newpage

\appendix
\renewcommand*{\thesection}{Appendix \Alph{section}}
\section{Tukey's Range Tests}\label{tukeys}

\begin{table}[h!]
	\centering
	\resizebox{\columnwidth}{!}{
	\begin{tabular}{@{}cccccc@{}}
		\toprule
		\textbf{GROUP1} & \textbf{GROUP2} &\textbf{ MEAN DIFF.} & \textbf{LOWER} & \textbf{UPPER} & \textbf{REJECT} \\ \midrule
		\texttt{MFR} & \texttt{MLPR} & 0.259 & 0.1317 & 0.3864 & True \\
		\texttt{MFR} & \texttt{MTR} & -0.0108 & -0.1382 & 0.1165 & False \\
		\texttt{MFR} & \texttt{PAR} & 0.2375 & 0.1102 & 0.3649 & True \\
		\texttt{MFR} & \texttt{RHAT} & 0.2461 & 0.1188 & 0.3735 & True \\
		\texttt{MFR} & \texttt{RHT} & 0.2546 & 0.1272 & 0.3819 & True \\
		\texttt{MFR} & \texttt{SGDR} & 0.2037 & 0.0764 & 0.3311 & True \\ \bottomrule
		\texttt{MLPR} & \texttt{MTR} & -0.2699 & -0.3972 & -0.1425 & True \\
		\texttt{MLPR} & \texttt{PAR} & -0.0215 & -0.1489 & 0.1058 & False \\
		\texttt{MLPR} & \texttt{RHAT} & -0.0129 & -0.1403 & 0.1144 & False \\
		\texttt{MLPR} & \texttt{RHT} & -0.0045 & -0.1318 & 0.1229 & False \\
		\texttt{MLPR} & \texttt{SGDR} & -0.0553 & -0.1827 & 0.072 & False \\ \bottomrule
		\texttt{MTR} & \texttt{PAR} & 0.2484 & 0.121 & 0.3757 & True \\
		\texttt{MTR} & \texttt{RHAT} & 0.257 & 0.1296 & 0.3843 & True \\
		\texttt{MTR} & \texttt{RHT} & 0.2654 & 0.138 & 0.3928 & True \\
		\texttt{MTR} & \texttt{SGDR} & 0.2146 & 0.0872 & 0.3419 & True \\ \bottomrule
		\texttt{PAR} & \texttt{RHAT} & 0.0086 & -0.1188 & 0.1359 & False \\
		\texttt{PAR} & \texttt{RHT} & 0.017 & -0.1103 & 0.1444 & False \\
		\texttt{PAR} & \texttt{SGDR} & -0.0338 & -0.1612 & 0.0936 & False \\ \bottomrule
		\texttt{RHAT} & \texttt{RHT} & 0.0084 & -0.1189 & 0.1358 & False \\
		\texttt{RHAT} & \texttt{SGDR} & -0.0424 & -0.1697 & 0.085 & False \\ \bottomrule
		\texttt{RHT} & \texttt{SGDR} & -0.0508 & -0.1782 & 0.0765 & False \\ \bottomrule
	\end{tabular}}
	\caption{Results of the Tukey's range test for experiment 1.}
	\label{exp1_tukey}
\end{table}

\newpage

\begin{table}[h!]
	\centering
	\resizebox{\columnwidth}{!}{	
	\begin{tabular}{@{}cccccc@{}}
		\toprule
		\textbf{GROUP1} & \textbf{GROUP2} &\textbf{ MEAN DIFF.} & \textbf{LOWER} & \textbf{UPPER} & \textbf{REJECT} \\ \midrule
		\texttt{MFR} & \texttt{MLPR} & -0.0223 & -0.1304 & 0.0859 & False \\
		\texttt{MFR} & \texttt{MTR} & -0.1953 & -0.3035 & -0.0871 & True \\
		\texttt{MFR} & \texttt{PAR} & -0.0381 & -0.1463 & 0.07 & False \\
		\texttt{MFR} & \texttt{RHAT} & -0.0428 & -0.151 & 0.0653 & False \\
		\texttt{MFR} & \texttt{RHT} & -0.0259 & -0.1341 & 0.0823 & False \\
		\texttt{MFR} & \texttt{SGDR} & -0.1408 & -0.249 & -0.0326 & True \\ \bottomrule
		\texttt{MLPR} & \texttt{MTR} & -0.173 & -0.2812 & -0.0649 & True \\
		\texttt{MLPR} & \texttt{PAR} & -0.0159 & -0.1241 & 0.0923 & False \\
		\texttt{MLPR} & \texttt{RHAT} & -0.0206 & -0.1287 & 0.0876 & False \\
		\texttt{MLPR} & \texttt{RHT} & -0.0036 & -0.1118 & 0.1045 & False \\
		\texttt{MLPR} & \texttt{SGDR} & -0.1185 & -0.2267 & -0.0103 & True \\ \bottomrule
		\texttt{MTR} & \texttt{PAR} & 0.1572 & 0.049 & 0.2653 & True \\
		\texttt{MTR} & \texttt{RHAT} & 0.1525 & 0.0443 & 0.2606 & True \\
		\texttt{MTR} & \texttt{RHT} & 0.1694 & 0.0612 & 0.2776 & True \\
		\texttt{MTR} & \texttt{SGDR} & 0.0545 & -0.0537 & 0.1627 & False \\ \bottomrule
		\texttt{PAR} & \texttt{RHAT} & -0.0047 & -0.1129 & 0.1035 & False \\
		\texttt{PAR} & \texttt{RHT} & 0.0122 & -0.0959 & 0.1204 & False \\
		\texttt{PAR} & \texttt{SGDR} & -0.1026 & -0.2108 & 0.0055 & False \\ \bottomrule
		\texttt{RHAT} & \texttt{RHT} & 0.0169 & -0.0913 & 0.1251 & False \\
		\texttt{RHAT} & \texttt{SGDR} & -0.0979 & -0.2061 & 0.0102 & False \\ \bottomrule
		\texttt{RHT} & \texttt{SGDR} & -0.1149 & -0.223 & -0.0067 & True \\ \bottomrule
	\end{tabular}}
	\caption{Results of the Tukey's range test for experiment 2.}
	\label{exp2_tukey}
\end{table}

\newpage

\begin{table}[h!]
	\centering
	\resizebox{\columnwidth}{!}{	
	\begin{tabular}{@{}cccccc@{}}
		\toprule
		\textbf{GROUP1} & \textbf{GROUP2} &\textbf{ MEAN DIFF.} & \textbf{LOWER} & \textbf{UPPER} & \textbf{REJECT} \\ \midrule
		\texttt{MFR} & \texttt{MLPR} & 0.2913 & 0.1947 & 0.3879 & True \\
		\texttt{MFR} & \texttt{MTR} & -0.0508 & -0.1474 & 0.0458 & False \\
		\texttt{MFR} & \texttt{PAR} & 0.2892 & 0.1926 & 0.3859 & True \\
		\texttt{MFR} & \texttt{RHAT} & 0.2955 & 0.1988 & 0.3921 & True \\
		\texttt{MFR} & \texttt{RHT} & 0.293 & 0.1963 & 0.3896 & True \\
		\texttt{MFR} & \texttt{SGDR} & 0.2874 & 0.1908 & 0.3841 & True \\ \bottomrule
		\texttt{MLPR} & \texttt{MTR} & -0.3421 & -0.4387 & -0.2455 & True \\
		\texttt{MLPR} & \texttt{PAR} & -0.0021 & -0.0987 & 0.0945 & False \\
		\texttt{MLPR} & \texttt{RHAT} & 0.0042 & -0.0925 & 0.1008 & False \\
		\texttt{MLPR} & \texttt{RHT} & 0.0016 & -0.095 & 0.0983 & False \\
		\texttt{MLPR} & \texttt{SGDR} & -0.0039 & -0.1005 & 0.0927 & False \\ \bottomrule
		\texttt{MTR} & \texttt{PAR} & 0.34 & 0.2434 & 0.4367 & True \\
		\texttt{MTR} & \texttt{RHAT} & 0.3463 & 0.2496 & 0.4429 & True \\
		\texttt{MTR} & \texttt{RHT} & 0.3438 & 0.2471 & 0.4404 & True \\
		\texttt{MTR} & \texttt{SGDR} & 0.3382 & 0.2416 & 0.4348 & True \\ \bottomrule
		\texttt{PAR} & \texttt{RHAT} & 0.0062 & -0.0904 & 0.1029 & False \\
		\texttt{PAR} & \texttt{RHT} & 0.0037 & -0.0929 & 0.1004 & False \\
		\texttt{PAR} & \texttt{SGDR} & -0.0018 & -0.0984 & 0.0948 & False \\ \bottomrule
		\texttt{RHAT} & \texttt{RHT} & -0.0025 & -0.0991 & 0.0941 & False \\
		\texttt{RHAT} & \texttt{SGDR} & -0.0081 & -0.1047 & 0.0886 & False \\ \bottomrule
		\texttt{RHT} & \texttt{SGDR} & -0.0055 & -0.1022 & 0.0911 & False \\ \bottomrule
	\end{tabular}}
	\caption{Results of the Tukey's range test for experiment 3.}
	\label{exp3_tukey}
\end{table}

\newpage

\begin{table}[h!]
	\centering
	\resizebox{\columnwidth}{!}{	
	\begin{tabular}{@{}cccccc@{}}
		\toprule
		\textbf{GROUP1} & \textbf{GROUP2} &\textbf{ MEAN DIFF.} & \textbf{LOWER} & \textbf{UPPER} & \textbf{REJECT} \\ \midrule
		\texttt{MFR} & \texttt{MLPR} & -0.0352 & -0.1058 & 0.0355 & False \\
		\texttt{MFR} & \texttt{MTR} & -0.2079 & -0.2786 & -0.1372 & True \\
		\texttt{MFR} & \texttt{PAR} & -0.0549 & -0.1255 & 0.0158 & False \\
		\texttt{MFR} & \texttt{RHAT} & -0.0494 & -0.1201 & 0.0212 & False \\
		\texttt{MFR} & \texttt{RHT} & -0.0307 & -0.1013 & 0.04 & False \\
		\texttt{MFR} & \texttt{SGDR} & -0.0604 & -0.131 & 0.0103 & False \\ \bottomrule
		\texttt{MLPR} & \texttt{MTR} & -0.1727 & -0.2434 & -0.1021 & True \\
		\texttt{MLPR} & \texttt{PAR} & -0.0197 & -0.0904 & 0.051 & False \\
		\texttt{MLPR} & \texttt{RHAT} & -0.0143 & -0.0849 & 0.0564 & False \\
		\texttt{MLPR} & \texttt{RHT} & 0.0045 & -0.0662 & 0.0752 & False \\
		\texttt{MLPR} & \texttt{SGDR} & -0.0252 & -0.0959 & 0.0455 & False \\ \bottomrule
		\texttt{MTR} & \texttt{PAR} & 0.153 & 0.0824 & 0.2237 & True \\
		\texttt{MTR} & \texttt{RHAT} & 0.1585 & 0.0878 & 0.2291 & True \\
		\texttt{MTR} & \texttt{RHT} & 0.1772 & 0.1066 & 0.2479 & True \\
		\texttt{MTR} & \texttt{SGDR} & 0.1475 & 0.0769 & 0.2182 & True \\ \bottomrule
		\texttt{PAR} & \texttt{RHAT} & 0.0054 & -0.0652 & 0.0761 & False \\
		\texttt{PAR} & \texttt{RHT} & 0.0242 & -0.0465 & 0.0949 & False \\
		\texttt{PAR} & \texttt{SGDR} & -0.0055 & -0.0762 & 0.0652 & False \\ \bottomrule
		\texttt{RHAT} & \texttt{RHT} & 0.0188 & -0.0519 & 0.0894 & False \\
		\texttt{RHAT} & \texttt{SGDR} & -0.0109 & -0.0816 & 0.0597 & False \\ \bottomrule
		\texttt{RHT} & \texttt{SGDR} & -0.0297 & -0.1004 & 0.0409 & False \\ \bottomrule
	\end{tabular}}
	\caption{Results of the Tukey's range test for experiment 4.}
	\label{exp4_tukey}
\end{table}

\newpage

\section*{References}

\bibliography{mybibfile}

\end{document}